\documentclass[twocolumn, switch]{article} %
\usepackage{datetime2}
\DTMsetdatestyle{ddmmyyyy}
\DTMsetup{datesep=/}
\usepackage{preprint}

\usepackage{amsmath, amsthm, amssymb, amsfonts}

\usepackage[
    style=nature,
    isbn=false,
    date=year,
    url=false,
    doi=false,
    maxbibnames=2,
]{biblatex}
\AtEveryBibitem{\clearlist{language}}
\AtEveryBibitem{\clearfield{eprint}}
\AtEveryBibitem{\clearfield{note}}
\AtBeginBibliography{\footnotesize}

\usepackage{url}

\usepackage[utf8]{inputenc}	%
\usepackage[T1]{fontenc}	%
\usepackage{xcolor}		%
\usepackage[colorlinks = true,
            linkcolor = purple,
            urlcolor  = blue,
            citecolor = cyan,
            anchorcolor = black]{hyperref}	%
\usepackage{booktabs} 		%
\usepackage{nicefrac}		%
\usepackage{microtype}		%
\usepackage{lineno}		%
\usepackage{float}			%

\usepackage{paralist}
\usepackage[title, page]{appendix}
\usepackage{placeins}
\usepackage{bbm}

\usepackage{caption}
\usepackage{newfloat}
\DeclareFloatingEnvironment[name={Supplementary Figure}]{suppfigure}
\usepackage{sidecap}
\sidecaptionvpos{figure}{c}

\usepackage{titlesec}
\titlespacing\section{0pt}{12pt plus 3pt minus 3pt}{1pt plus 1pt minus 1pt}
\titlespacing\subsection{0pt}{10pt plus 3pt minus 3pt}{1pt plus 1pt minus 1pt}
\titlespacing\subsubsection{0pt}{8pt plus 3pt minus 3pt}{1pt plus 1pt minus 1pt}

\usepackage{tikz,xcolor,hyperref}

\definecolor{lime}{HTML}{A6CE39}
\DeclareRobustCommand{\orcidicon}{
	\begin{tikzpicture}
	\draw[lime, fill=lime] (0,0) 
	circle [radius=0.16] 
	node[white] {{\fontfamily{qag}\selectfont \tiny ID}};
	\draw[white, fill=white] (-0.0625,0.095) 
	circle [radius=0.007];
	\end{tikzpicture}
	\hspace{-2mm}
}
\foreach \x in {A, ..., Z}{\expandafter\xdef\csname orcid\x\endcsname{\noexpand\href{https://orcid.org/\csname orcidauthor\x\endcsname}
			{\noexpand\orcidicon}}
}
\usepackage{pgfplots}
\usetikzlibrary{positioning}

\tikzstyle{causalNode} = [circle, draw, thick, minimum size=2.25em]
\tikzstyle{cause} = [ ->, ultra thick]
\tikzstyle{causeDeterministic} = [->, ultra thick, dash dot]
\tikzstyle{causeMissingData} = [->, ultra thick, loosely dashed, red]

\newcommand{\refArrow}[1]{$\begin{tikzpicture}[baseline=-0.63ex]%
\node (A) {};%
\node (B) [right of=A, node distance=3em] {};%
\draw[#1] (A) -- (B);%
\end{tikzpicture}$}

\newcommand{\EE}[0]{\mathbb{E}}
\newcommand{\ddo}{\mathrm{do}}

\newcommand{\singleColWidth}[0]{88mm}
\newcommand{\doubleColWidth}[0]{180mm}

\title{Evaluating vaccine allocation strategies using simulation-assisted causal modelling}
\date{\DTMdisplaydate{2022}{12}{14}{-1}}

\usepackage{authblk}

\author[1]{Armin~Keki\'c\orcidA{}}
\author[2]{Jonas~Dehning\orcidB{}}
\author[1]{Luigi~Gresele\orcidC{}}
\author[1,3]{Julius~von~Kügelgen\orcidD{}}
\author[2,4,*]{\\ Viola~Priesemann\orcidE{}}
\author[1,*]{Bernhard~Schölkopf\orcidF{}}
\affil[1]{Max Planck Institute for Intelligent Systems, Tübingen, Germany}
\affil[2]{Max Planck Institute for Dynamics and Self-Organization, Göttingen, Germany}
\affil[3]{Department of Engineering, University of Cambridge, United Kingdom}
\affil[4]{Georg August University, Göttingen, Germany}
\affil[*]{Joint supervision}
\begin{document}

\twocolumn[ %
  \begin{@twocolumnfalse} %
  
\maketitle

\begin{abstract}
\noindent 
\looseness-1 Early on during a pandemic, vaccine availability is limited, requiring prioritisation of different population groups.
Evaluating vaccine allocation is therefore a crucial element of pandemics response.
In the present work, we develop a model to retrospectively evaluate age-dependent counterfactual vaccine allocation strategies against the COVID-19 pandemic. 
To estimate the effect of allocation on the expected severe-case incidence, we employ a simulation-assisted causal modelling approach which combines a compartmental infection-dynamics simulation, a coarse-grained, data-driven causal model and literature estimates for immunity waning.
We compare Israel's implemented vaccine allocation strategy in 2021 to counterfactual strategies such as no prioritisation, prioritisation of younger age groups or a strict risk-ranked approach;
we find that Israel's implemented strategy was indeed highly effective.
We also study the marginal impact of increasing vaccine uptake for a given age group and find that increasing vaccinations in the elderly is most effective at preventing severe cases, whereas additional vaccinations for middle-aged groups reduce infections most effectively.
Due to its modular structure, our model can easily be  adapted to study future pandemics.
We demonstrate this flexibility by investigating vaccine allocation strategies for a pandemic with characteristics of the Spanish Flu.
Our approach thus helps evaluate vaccination strategies under the complex interplay of core epidemic factors, including age-dependent risk profiles, immunity waning, vaccine availability and spreading rates.
\end{abstract}
\vspace{0.35cm}

  \end{@twocolumnfalse} %
] %

\section{Introduction} %
\label{sec:introduction}
\begin{figure*}[htb]
    \centerline{
        \includegraphics[width=18cm]{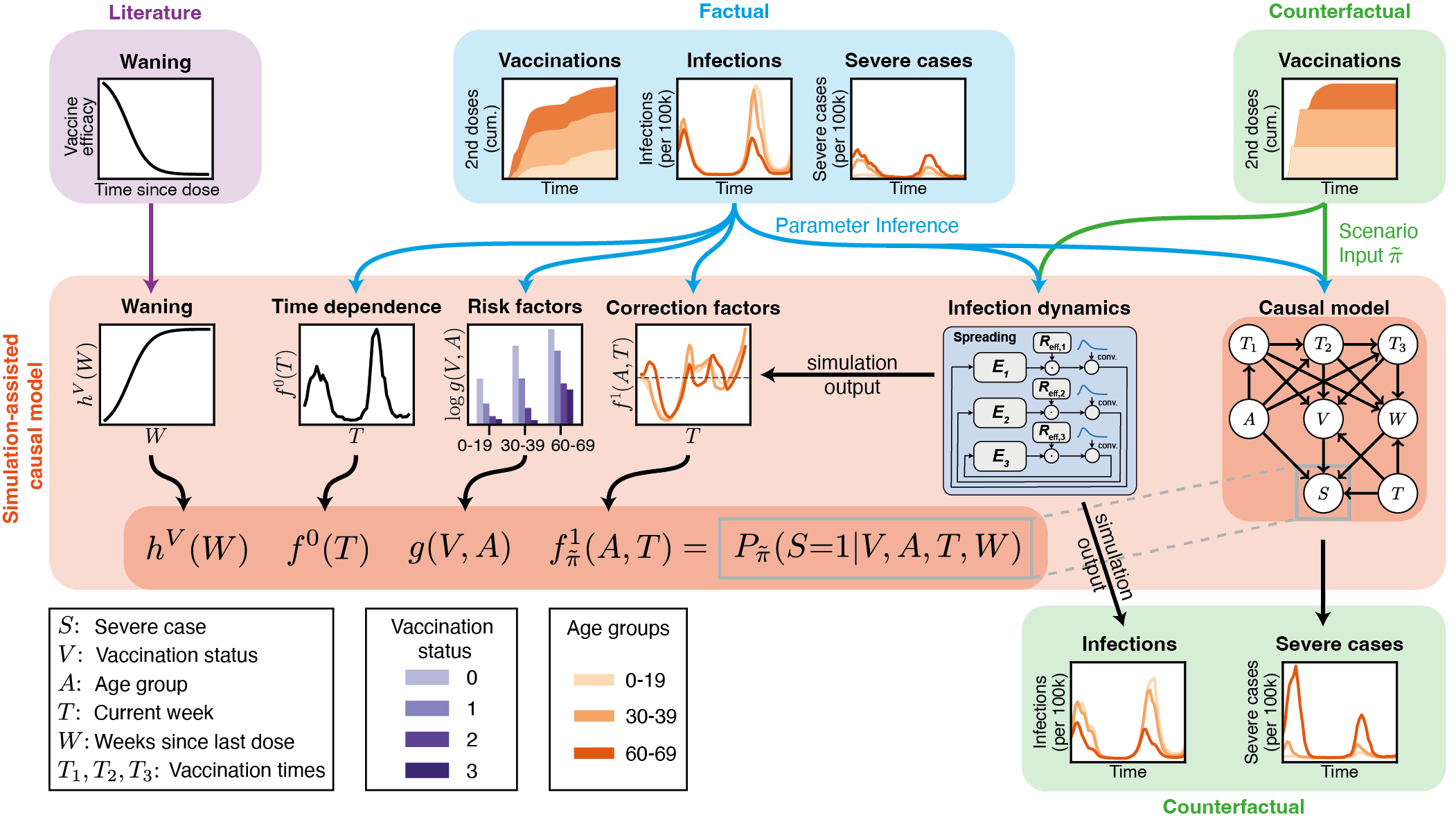}
    }
    \caption{
    \textbf{Method overview.}
    Our goal is to compute the weekly severe-case incidence (bottom right) under counterfactual vaccine allocation strategies (in the example shown here: \texttt{YoungFirst}, top right).
    To compute this counterfactual scenario, we provide an estimate of the severity mechanism $P_{\tilde \pi}(S{=}1|V, A, T, W)$ for all combinations of $(V, A, T, W)$ through our proposed factorisation~\eqref{eqn:severity_factorisation}.
    The other conditionals in the causal graph are directly estimated from data (factual strategy, top centre) or are intervened upon according to the counterfactual strategy.
    The risk factors $g(V,A)$ and the time dependence $f^0(T)$ are estimated from data after accounting for immunity waning~$h(W)$ derived from literature estimates (top left).
    An SEIR-like infection dynamics model is fit to the factual infections and subsequently used to simulate infections under the counterfactual strategy. 
    The simulation output is used to compute the correction factors $f^1_{\tilde \pi}(A, T)$ accounting for the age-specific change in probability of being infected.
    Only three of nine age groups are shown for simplicity.
    }
    \label{fig:methods_diagram}
\end{figure*}%
\noindent The COVID-19 pandemic posed significant challenges to societies and decision makers around the world.
Many governments implemented non-pharmaceutical interventions to limit the spread of infections and reduce the number of severe cases~\cite{brauner_inferring_2020, dehning_inferring_2020}.
The development of efficient vaccines has provided another key control measure to combat the COVID-19 pandemic~\cite{OliuBarton2022}.
However, vaccine supply can fail to meet demand, and vaccine uptake can fall short of expectations.
Under these conditions, governments have to find rational strategies to allocate vaccines to minimise harm.
It is therefore important to understand how to evaluate and compare different vaccine allocation strategies. 
\par
One crucial aspect to consider when designing such strategies is age, which is a key risk factor for COVID-19 mortality~\cite{Odriscoll2021}.
While pre-existing conditions and the high exposure of health care workers have also played a role in vaccine prioritisation~\cite{Rosen2021}, in this work, we focus on age-dependence, including interactions between age groups, as one of the most important factors.
Given an observed evolution of infections and severe cases, we seek to answer central policy questions:
\emph{Given limited vaccine availability, should one have prioritised first vaccinations of the young or booster shots for the elderly?}
\emph{Which age groups should have been targeted preferentially to increase vaccine uptake?}
\emph{How would a different age-dependent risk profile have impacted the outcome of a vaccination policy?}
Answering such questions requires computing the effects of hypothetical interventions on a complex system (given observations of the same system under different conditions).
Such \textit{What if?}-scenarios lie at the heart of causal inference and relate to \emph{counterfactual} reasoning: that is, reasoning about how events would have turned out had some circumstances been altered~\cite{Pearl2009,rubin1974estimating}.
\par
The gold standard for inferring average causal effects are randomised controlled trials (RCTs)~\cite{Fisher1935}, which are used to evaluate medical treatments such as COVID-19 vaccines~\cite{Polack2020}.
However, to compare full vaccine allocation strategies at the country level, running an RCT is infeasible in practice (there is only one copy of each country), mirroring similar challenges in the empirical estimation of individualised treatment effects~\cite{holland1986statistics,shalit2017estimating}.
Moreover, it would be ethically unacceptable to implement vaccine allocation strategies that are expected to be suboptimal for the population.
Here, modelling approaches provide an important tool to fill this gap. 
For COVID-19, one and a half years into the vaccination campaign, we now have data (age-resolved cases, hospitalisations and vaccination times) to infer the reduction of spread due to vaccination across different age groups and subsequently simulate counterfactual vaccination scenarios.
\par 
\looseness=-1 To capture the effects of changes in vaccine allocation strategy, we have to model their impact on spread and hospitalisation.
Furthermore, we need to consider aspects such as vaccine efficacy, immunity waning, age-dependent risk profiles and contact structures.
Two established modelling paradigms are compartmental differential equation models and machine learning (ML) approaches.
\par 
In principle, compartmental models like the susceptible-infectious-recovered (SIR) model (and its extensions) can be used to answer the types of questions we are interested in---provided that all relevant parameters are known sufficiently well~\cite{Sunohara2021, Tuite2010, foy_comparing_2021, Han2021, bauer2021relaxing}.
However, this is typically not the case. 
Thus, a framework combining inference of parameters and prediction is necessary.
To jointly model infections and severe cases, compartmental models require a large state space, whose parameters can be difficult to estimate from data without overfitting~\cite{Pollicott2012}.
\par
\looseness=-1 ML methods excel at fitting data and making predictions based on statistical associations, but are generally unable to answer causal questions.
Moreover, they are unreliable when the underlying data distribution changes. 
Yet, we are precisely interested in how our system behaves under distribution shifts: we want to know how the expected severe-case incidence would have changed had we implemented different vaccine allocation strategies.
\par
\looseness=-1 Causal models occupy a middle ground between the two paradigms and are better-suited for our purpose since they are \emph{modular} and \emph{interventional}.
Modularity refers to a model being composed of autonomous components or subsystems and allows for combining different sources of knowledge.
Interventions are naturally supported by causal models since they explicitly capture the data-generating mechanisms rather than mere statistical associations.
Causal models can thus answer counterfactual questions while clearly expressing the underlying assumptions~\cite{Pearl2009, peters2017elements}.
However, using them to describe time-varying systems and modelling the dynamics of epidemic spread is cumbersome.
\par
Since neither compartmental nor causal models on their own are fully suited for our task, we resort to a hybrid modelling approach involving a modular combination thereof.
We propose a coarse-grained causal model in which most components are estimated from data, but where additionally one of the modules is a compartmental model and another one is derived from literature estimates.
In particular, we use a susceptible-exposed-infectious-recovered (SEIR) model for the infection dynamics---but not for severe cases---and rely on literature estimates of immunity waning, as illustrated in Fig.~\ref{fig:methods_diagram}.
The SEIR-like infection-dynamics model can be fit more easily to data compared to a joint model of severe cases and infections.
For severe cases, given a set of qualitative causal assumptions, fitting the causal model reduces to the problem of statistical estimation of conditional probabilities.   
This combines the strengths of the two approaches: the data-driven nature of causal models and the expressivity of compartmental models.
\par
As a case study, we apply our method to a comprehensive dataset collected in Israel~\cite{Israel2021}.
Specifically, we compare several counterfactual age-dependent vaccine allocation strategies to the factual strategy, assuming a fixed number of administered doses and fixed vaccine uptake rate per age group.
We also simulate the effect of campaigning for vaccine uptake in a given age group by increasing vaccine uptake rate in one group and computing the effect on the severe-case incidence across all age groups.
To showcase the capability of our model to change and examine the influence of individual modules, we consider a different type of disease whose age-dependent risk profile is based on the Spanish Flu.
We also investigate the effect of waning immunity by changing the timescale at which immunity weakens.
\section{Results} %
\label{sec:results}

\subsection{Methods summary} %
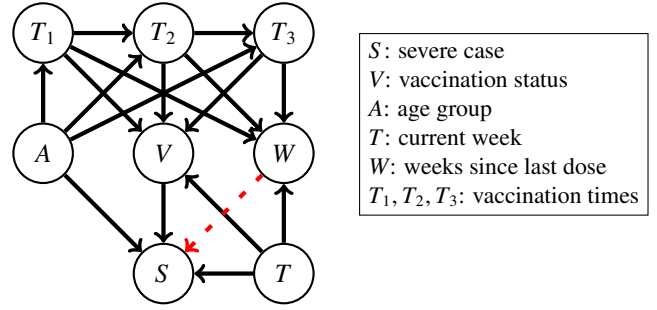
\begin{figure}[htb]
    \centering
    \newcommand{\pos}{1.6}
\begin{tikzpicture}[causal_node/.style={circle, draw, thick, minimum size=2.25em}]
    \node[causal_node] (t3) at (\pos,\pos){$T_3$};
    \node[causal_node] (t2) at (0,\pos){$T_2$};
    \node[causal_node] (t1) at (-\pos,\pos){$T_1$};

    \node[causal_node] (a) at (-\pos,0){$A$};
    \node[causal_node] (v) at (0,0){$V$};
    \node[causal_node] (w) at (\pos,0){$W$};
    
    \node[causal_node] (s) at (0,-\pos){$S$};
    \node[causal_node] (t) at (\pos,-\pos){$T$};
    
    \draw[cause] (t1) -- (t2);
    \draw[cause] (t2) -- (t3);

    \draw[cause] (a) -- (t1);
    \draw[cause] (a) -- (t2);
    \draw[cause] (a) -- (t3);
    
    \draw[cause] (t1) -- (v);
    \draw[cause] (t2) -- (v);
    \draw[cause] (t3) -- (v);
    \draw[cause] (t) -- (v);

    \draw[cause] (t1) -- (w);
    \draw[cause] (t2) -- (w);
    \draw[cause] (t3) -- (w);
    \draw[cause] (t) -- (w);

    \draw[cause] (t) -- (s);
    \draw[cause] (a) -- (s);
    \draw[cause] (v) -- (s);
    \draw[causeMissingData] (w) -- (s);
    
    \node[draw,align=left, right of = t3, anchor=north west] {
    \footnotesize $S$: severe case\\ 
    \footnotesize $V$: vaccination status\\ 
    \footnotesize $A$: age group\\ 
    \footnotesize $T$: current week\\ 
    \footnotesize $W$: weeks since last dose \\
    \footnotesize $T_1, T_2, T_3$: vaccination times
    };
\end{tikzpicture}
    \caption[
    	\textbf{Causal graph used to model the variables influencing severe Covid-19 cases $S$.}
    	We consider the variables vaccination status $V$, age group $A$, the current week $T$ and the time since the last dose was administered $W$.
    	$V$ and $W$ deterministically depend on the current week $T$ and the vaccination times $T_1, T_2, T_3$ for each dose.
    	An arrow indicates that one variable has a direct causal influence on another.
    	The red dashed arrow indicates a relationship which cannot be estimated because the data is incomplete.
    ]{
    	\textbf{Causal graph used to model the variables influencing severe Covid-19 cases $S$.}
    	We consider the variables vaccination status $V$, age group $A$, the current week $T$ and the time since the last dose was administered $W$.
    	$V$ and $W$ deterministically depend on the current week $T$ and the vaccination times $T_1, T_2, T_3$ for each dose.
    	An arrow indicates that one variable has a direct causal influence on another.
    	The red dashed arrow (\refArrow{causeMissingData}) indicates a relationship which cannot be estimated because the data is incomplete.
    }
    \label{fig:causal_graph}
\end{figure}

\label{sub:methods_summary}
\noindent We use a causal graphical model~\cite{Pearl2009}, as shown in Figure~\ref{fig:causal_graph}, to describe an individual's probability of developing severe COVID-19.
We use a binary variable $S\in\{0, 1\}$ where $1$ indicates a severe case and $0$ describes a mild case or no infection at all.
We assume the severe-case probability depends on the following variables: 
the vaccination status of the individual $V\in \{0,1,2,3\}$ indicating the number of vaccine doses a person has received;
their age group $A \in \{\text{0-19}, \text{20-29}, \dots, \text{80-89}, \text{90+}\}$; 
the current week $T\in \{1, \dots, M\}$; 
and the waning time, \textit{i.e.}\ the time since the last dose was received $W\in \{1, \dots, M\}$.
$M$ is the number of weeks in the considered time window.
The variables $V$ and $W$ are functions of the weeks in which the respective doses were received $T_i\in \{1, \dots, M{+}1\}$ for $i\in\{1, 2, 3\}$ and the current week $T$.
By common convention, we denote random variables by uppercase letters and realisations thereof by lowercase letters.
\par
We are interested in how interventions on the distribution of vaccination times for different age groups $A$ affect the expected severe-case probability:
\begin{equation}
	\EE[S|\ddo(T_1, T_2, T_3 \sim \tilde P(T_1, T_2, T_3|A) )]
	\label{eqn:severity_post_intervention}
\end{equation}
where we denote the distributions of vaccination times pre- and post-intervention as $\pi = P(T_1, T_2, T_3|A)$ and $\tilde \pi = \tilde P(T_1, T_2, T_3|A)$, respectively.
The $\ddo(\cdot)$ operator describes a change in distribution arising from an intervention~\cite{Pearl2009} (see Section~\ref{sub:target_function} for details).
\par
There are three challenges we have to address: 
\begin{inparaenum}[(i)]
    \item In the publicly available data by Israel's Ministry of Health~\cite{Israel2021} severe outcomes are not registered as a function of time since the last dose was received.
	We only have access to the marginal distribution $P(S|V, A, T) = \sum_w P(S|V, A, T, w) P(w|V, A, T)$.
	\item Computing the post-intervention severity \eqref{eqn:severity_post_intervention} involves evaluating the conditional $P(S|V, A, T)$ for combinations $(V, A, T)$ for which there are no observations.
	In particular, due to the implemented age-ranked vaccine allocation strategy in Israel, there may not have been any vaccinated subjects in certain younger age groups for some of the early weeks.
	\item Changing the vaccine allocation strategy influences the probability of having a severe case in two ways: firstly, by changing the probability of having immunity through vaccination and secondly, through impacting the infection dynamics at the population level. 
	Such changes in infection dynamics are not captured by the causal model alone.
\end{inparaenum}
\par
\begin{figure}[htb]
    \centerline{
        \includegraphics[width=\singleColWidth]{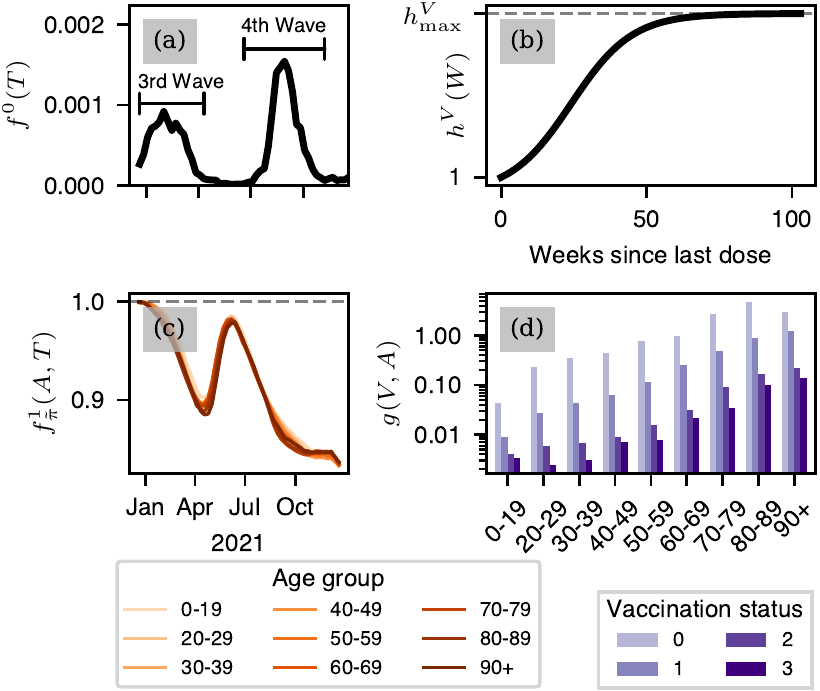}
    }
    \caption{
    \textbf{Factors contributing to the severe case probability.}
    (a) Estimated time dependence $f^0(T)$ approximately following the two infection waves in Israel in 2021.
    (b) The waning curve $h^V(W)$ with increasing risk over time as the immunity wanes.
    The waning curve is computed based on results from~\cite{tartof2021}.
    (c) Estimated age-dependent correction factors $f^1_{\tilde \pi}(A, T)$ accounting for the change in population-level infection dynamics.
    The correction factors shown here correspond to the scenario with increased vaccine uptake rate by 2\%, leading to a relative decrease in the number of infections.
    (d) Risk factor $g(V, A)$ estimates indicating the relative risk of having a severe case by age $A$ and vaccination status $V$.
    }
    \label{fig:severity_factorisation}
\end{figure}
To address these challenges, we propose a factorisation of the \emph{severity mechanism}
\begin{equation}
	P_{\tilde \pi}(S{=}1|V, A, T, W) = f^0(T)\  g(V, A)\ h^V(W)\ f^1_{\tilde \pi}(A, T). \label{eqn:severity_factorisation}
\end{equation}
The observed aggregate time dependence of the probability of having a severe case is captured by $f^0(T)$. 
The factor $g(V, A)$ describes the age- and vaccination-status-dependent relative risk factor of having a severe case, where we normalize $g=1$ for the unvaccinated 60-69 year-olds.
The factor $h^V(W)$ describes the waning of immunity against infection.
Finally, $f^1_{\tilde  \pi}(A,T)$ is a correction factor that depends on the post-intervention vaccination distribution and accounts for the change in infection dynamics.
The subscript $\tilde \pi$ indicates factors that depend on the post-intervention vaccine allocation strategy.
\par
\looseness=-1 The factorisation~\eqref{eqn:severity_factorisation} resolves the challenges above by allowing us to
\begin{inparaenum}[(i)]
	\item incorporate literature knowledge about immunity waning into our causal model,
	\item estimate $P(S{=}1|V, A, T, W)$ for all values in the conditioning set by transferring knowledge between vaccination states, age groups and weeks, and
	\item take into account the population level impact on the infection dynamics.
\end{inparaenum}
\par
$h^V(W)$ can be derived from literature estimates for the vaccine efficacy against infection as a function of time since the last dose was administered~\cite{tartof2021}.
$f^0(T)$ and $g(V,A)$ can be estimated from data after correcting for the influence of waning.
The correction factor $f^1_{\tilde  \pi}(A,T)$ under the counterfactual vaccine allocation strategy is given by the relative change in weekly infection probability for each age group
\begin{equation}
   f^1_{\tilde  \pi}(A,T)  = \frac{P_{\tilde \pi}(I{=}1|V, A, T, W)}{P_{\pi}(I{=}1|V, A, T, W)} . \label{eqn:correction_factor}
\end{equation}
Hence, the correction factor couples the compartmental model for infection dynamics with the causal model.
The derivations of all estimators are given in Section~\ref{sub:estimating_the_severity_conditional_factors}.
Estimated factors are shown in Figure~\ref{fig:severity_factorisation}.
\par
\begin{figure}[htb]
    \centering
    \includegraphics[width=\singleColWidth]{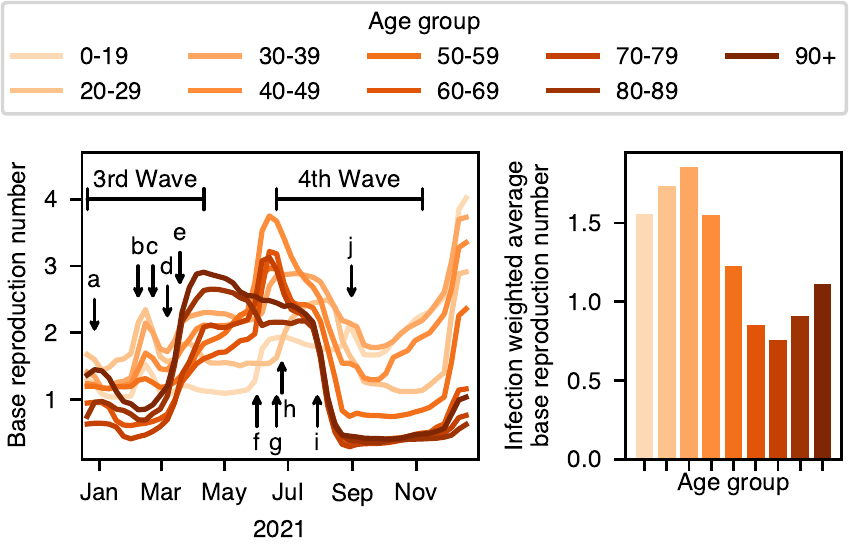}
    \caption{
    \textbf{Base reproduction numbers per age group inferred by the SEIR-like infection dynamics model.}
    The base reproduction numbers describe the contribution of each age group to infection spread after accounting for the effect of vaccination.
    Left: inferred weekly base reproduction numbers between \DTMdisplaydate{2020}{12}{20}{-1} and \DTMdisplaydate{2021}{12}{25}{-1};
    (a): lockdown~\cite{Jerusalem2020},
    (b-e): restriction easing phases 1-4~\cite{katz2021lessons},
    (f): Green Pass and Purple Badge requirement lifted,
    (g): high and middle school summer break start, 
    (h): indoor mask requirements~\cite{Toi2021},
    (i): Green Pass requirements~\cite{Jerusalem2021},
    (j): school summer break end.
    The ticks on the x-axis indicate the first day of the respective month.
    Right: average base reproduction numbers weighted by the factual total weekly infections.
    }
    \label{fig:reproduction_number}
\end{figure}

To estimate the effect of changing the vaccine allocation strategy on the infection dynamics ($P_{\tilde \pi}(I{=}1|V, A, T, W)$), we first infer the parameters of a Bayesian SEIR-like model under the factual strategy $\pi$.
Vaccines are assumed to offer some protection against infection: $70\,\%$, $90\,\%$ and $95\,\%$ directly after the first, second and third dose, respectively; after which the protection is waning at the same rate~\cite{tartof2021}. 
For each age group, we fit a time-dependent \emph{base reproduction number}, that is, the reproduction number in a hypothetical non-immune population (Fig.~\ref{fig:reproduction_number} left).
We assume a generation interval of 4 days \cite{pung_serial_2021, hart_generation_2022} and a reporting delay of 6 days. 
Non-equal reproduction numbers for every age group are modeled by modulating symmetrically the rows and columns of a contact matrix.
The preference for contacts within each age group is parameterised by a contact mixing factor $\gamma$ between 0 (no mixing between age groups) and 1 (all-to-all connections).
By default, this factor is set to 0.8; we show that results are similar with lower and higher mixing factors in Supplement~\ref{app:results_assuming_other_mixing_factors}.
With the inferred reproduction number we rerun the model with the counterfactual vaccine allocation strategy $\tilde \pi$ to obtain an estimate of the number of infections in the counterfactual scenario. The simulated number of infections is then used to calculate the correction factor~\eqref{eqn:correction_factor}.

\subsection{Counterfactual vaccine allocation strategies} %
\label{sub:counterfactual_vaccine_allocation_strategies}
\begin{figure}[htb]
    \centering
    \includegraphics[width=\singleColWidth]{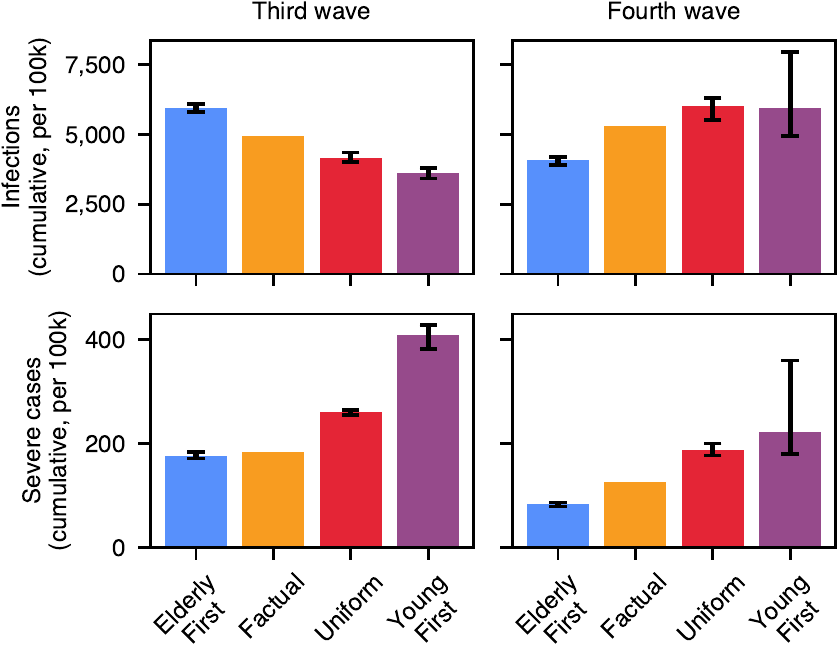}
    \caption{
    \looseness=-1
    \textbf{Cumulative incidences of infections (top row) and severe cases (bottom row) for the two infection waves in 2021 under the factual and counterfactual vaccine allocation strategies.}
    For the third wave we sum all cases from \DTMdisplaydate{2020}{12}{20}{-1} to \DTMdisplaydate{2021}{04}{11}{-1}; for the fourth wave from \DTMdisplaydate{2021}{06}{20}{-1} to \DTMdisplaydate{2021}{11}{07}{-1}.
    The whiskers show the $95\,\%$ credible intervals.
    }
    \label{fig:policy_exp_cumulative}
\end{figure}
\begin{figure*}[htb]
    \centering
    \centerline{
        \includegraphics[width=\doubleColWidth]{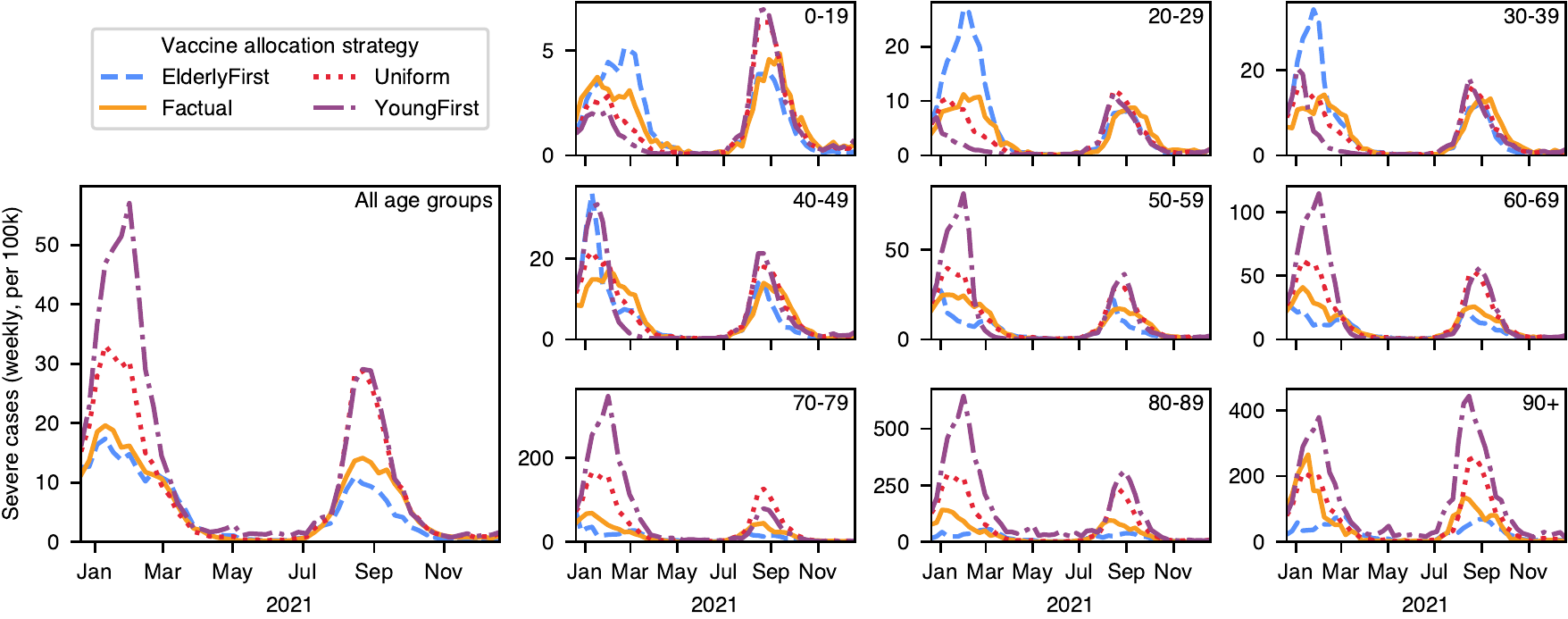}
    }
    \caption{
    \textbf{Expected weekly number of severe-case incidence for the entire population (left) and in each age group (right) for the factual and counterfactual vaccine allocation strategies.}
    The right panels show the trade-offs in severe-case incidence between age groups under different vaccine allocation strategies.
    The ticks on the x-axis indicate the first day of the respective month.
    }
    \label{fig:policy_exp_overview}
\end{figure*}
We compare four age-dependent vaccine allocation strategies:
the factual and three counterfactual ones.
\begin{inparaenum}[(i)]
    \item \texttt{Factual}: The vaccine allocation strategy implemented in Israel generally prioritised the elderly (starting with all people aged $60$ and over) for both the initial two  doses and for booster shots, but also prioritised nursing home residents, patients with pre-existing medical conditions, and front-line health care workers~\cite{Rosen2021}.
    \item \texttt{Uniform}: In the uniform strategy, we do not implement any prioritisation based on age or any other factors; all age groups are vaccinated at the same rate.
    \item \texttt{ElderlyFirst}: We prioritise age groups in descending order, starting with the oldest and ending with the youngest.
    This strategy differs from the factual strategy in that it strictly prioritises by age and does not consider other factors.
    \item \texttt{YoungFirst}: the opposite of \texttt{ElderlyFirst}.
\end{inparaenum}
\paragraph{Infections}  %
As shown in Figure~\ref{fig:policy_exp_cumulative}, during the third wave, \texttt{YoungFirst} leads to the lowest infection incidence, followed by \texttt{Uniform} and the \texttt{Factual} strategy.
\texttt{ElderlyFirst} leads to the highest infection incidence.
\par %
The most effective strategy in preventing infections is determined by which age group is contributing most to infection spread.
The \emph{base} reproduction numbers express how much an age group contributes to infection spread after removing the effect of vaccinations; in other words, how much an age group would contribute to infection spread if no one in that age group were vaccinated.
As shown in Figure~\ref{fig:reproduction_number}, during the third wave, the base reproduction numbers tended to be higher among the young and middle age groups and lower for the elderly, pointing to differences in behaviour and average number of contacts.
Therefore, strategies that prioritise young and middle age groups are most effective at preventing infections.
\par %
\looseness=-1 During the fourth wave, the relative effectiveness of strategies at preventing infections is reversed. 
\texttt{ElderlyFirst} leads to the lowest infection incidence, followed by the \texttt{Factual} strategy.
\texttt{YoungFirst} and \texttt{Uniform} lead to the most number of infections.
\par %
The order of the vaccine allocation strategies regarding infections in the fourth wave is changed since the estimated base reproduction numbers per age group are different.
For the first half of the wave, the base reproduction number in the youngest age group, accounting for roughly a third of the population, tends to be lower than for the other age groups, presumably due to the school summer break, as indicated in Figure~\ref{fig:reproduction_number}.
Conversely, at the start of the wave, the base reproduction numbers for the middle age groups and the elderly tend to be higher.
Therefore, prioritising these age groups is a more effective measure for preventing infection spread at that point in time.

\paragraph{Severe cases}  %
\looseness=-1 As shown in Figure~\ref{fig:policy_exp_cumulative}, the \texttt{ElderlyFirst} strategy leads to the lowest cumulative \emph{severe-case} incidence: 
it performs similar (third wave) or better than (fourth wave) the \texttt{Factual} strategy.
The \texttt{Uniform} and \texttt{YoungFirst} strategies lead to the highest cumulative severe-case incidence.
Which one of these two leads to the highest severe-case incidence depends on the wave and the assumed contact mixing factor, see Supplement~\ref{app:results_assuming_other_mixing_factors}.
\par 
Figure~\ref{fig:policy_exp_overview} shows the trade-offs made between age groups in terms of severe cases under the different strategies.
As expected, the younger age groups benefit most from the \texttt{YoungFirst} strategy and older age groups experience the lowest severe-case incidence under the \texttt{ElderlyFirst} strategy.
\par 
The severe-case risk is lowest for the youngest age groups, as shown in Figure~\ref{fig:severity_factorisation}(d).
Hence, prioritising the younger age groups leaves the older age groups (which are at higher risk) less protected against severe cases.
While the \texttt{YoungFirst} strategy reduces the severe-case incidence in the youngest age group, this reduction is far outweighed by the increase in most other age groups.
We find the opposite effect for the \texttt{ElderlyFirst} scenario: we have a reduction of the severe-case incidence in the elderly and an increase in the younger age groups.
\par 
Our results suggest that under the \texttt{ElderlyFirst} strategy, the cumulative severe-case incidence could have been reduced had this strategy been implemented in Israel ($177$ vs.\ $184$ per 100k and $84$ vs.\ $126$ per 100k in the third and fourth waves, respectively).
However, our counterfactual vaccine allocation strategies make some simplifying assumptions that are difficult to implement in practice.
We assume that it is possible to vaccinate all willing patients of an age group before moving on to the next age group without delays.
In practice, this is difficult to accomplish, in particular given that it may be harder to reach the elderly.
Therefore, no realistic vaccine allocation strategy can be as strict as the protocol followed in \texttt{ElderlyFirst} and will have some overlap between age groups.
Hence, from our counterfactual results and these observations, we conclude that the \texttt{Factual} strategy may have been close to optimal.

\subsection{Impact of increasing vaccine uptake} %
\label{sub:impact_of_increasing_vaccine_acceptance_rate}
\begin{figure}[t!]
    \centering
    \includegraphics[width=\singleColWidth]{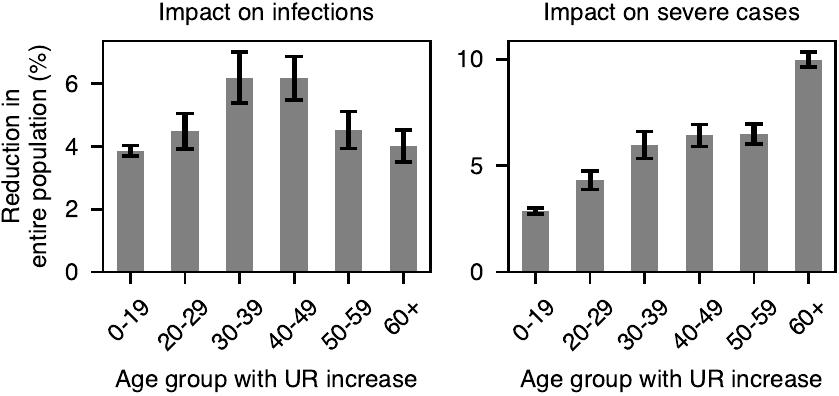}
    \caption{
        \textbf{Impact of increasing vaccine uptake rate (UR) in a given age group.}
        In each scenario, the vaccine uptake rate is increased in a given age group by a fixed number corresponding to $0.6\,\%$ of the population being motivated to get vaccinated.
        We assume that the change comes from originally unvaccinated individuals who are persuaded to receive three doses.
        The plots show the impact on cumulative infections (left) and severe cases (right) in the entire population---not just in the age group in which the UR was increased.
        We consider cases from \DTMdisplaydate{2020}{12}{20}{-1} to \DTMdisplaydate{2021}{12}{25}{-1}.
        While the middle age groups have a larger impact on the infection dynamics, this effect is outweighed by the higher severe-case risk in the $60+$ age group when considering the impact on severe cases.
        The whiskers show the $95\,\%$ credible intervals.
        }
    \label{fig:acceptance_exp_overview}
\end{figure}
We simulate the effect of increasing vaccine uptake in a single age group.
Limited vaccine uptake is a factor that prolongs the necessity of non-pharmaceutical interventions~\cite{OliuBarton2022}.
One possible intervention governments have at their disposal is to encourage vaccinations through advertisement campaigns.
Such campaigns can be targeted at specific age groups by choosing the channel over which the campaign is broadcast.
Our method allows us to estimate the impact of increasing vaccine uptake on the expected incidence of infections and severe outcomes.
We simulate the effect of increasing the vaccine uptake rate within one \textit{single given age group} by administering an additional $N{=}55, 746$ doses ($0.6\,\%$ of the total population) to that age group.
All other age groups follow the factual vaccine allocation strategy.
The additional doses are spread over the entire time period by scaling the weekly administered doses by a constant factor.
\par %
Increasing vaccine uptake in the oldest age groups is most effective in decreasing the severe-case incidence in the total population, as shown in Figure~\ref{fig:acceptance_exp_overview}.
On the other hand, increasing vaccine uptake in the middle age groups is most effective in decreasing the infection incidence in the total population.
\par %
There are two effects of an increase in vaccine uptake.
Firstly, there is a larger number of individuals who are better protected against infections and severe cases through vaccine-induced immunity.
This effect can be explained by the risk factors shown in Figure~\ref{fig:severity_factorisation}(d): by increasing the vaccine uptake rate in a given age group, we effectively move an additional part of this population from vaccination status 0 (unvaccinated) to 3 (boostered).
We find the largest effect for the elderly, since they have the largest absolute difference in risk factors between vaccination states 0 and 3.
Secondly, there is a population-level effect: by influencing the infection spread, the total number of infections is reduced (see~Figure~\ref{fig:acceptance_exp_overview} (left)).
Since the base reproduction number tends to be higher for the middle age groups (see~Figure~\ref{fig:reproduction_number} (right)) they have a larger impact on the number of infections when their vaccine uptake rate is increased.
However, when considering the severe-case incidence, this effect on infections is not large enough to outweigh the differences in risk factors, as shown in Figure~\ref{fig:acceptance_exp_overview} (right).
In summary, even when taking the effect on infection dynamics into account, it would have been most beneficial to increase vaccine uptake rate in the elderly.

\subsection{Simulating other disease types} %
\label{sub:simulating_other_disease_types}
To investigate if one can generalise the recommendation to first vaccinate age groups with the highest severe-case risk, we simulate different types of diseases by adopting other risk factors.
The explicit factorisation of the severity mechanism~\eqref{eqn:severity_factorisation} allows us to dissect the different contributing factors that determine the expected severe-case incidence.
One of those factors is the age- and vaccination-dependent risk profile of COVID-19.
Here, we showcase the ability of our model to be adapted to different diseases in fictional, yet plausible scenarios.
We compare three disease types, shown in Figure~\ref{fig:risk_profile_exp} (left):
\begin{inparaenum}[(i)]
    \item COVID-19 (see Figure~\ref{fig:severity_factorisation}).
    \item Spanish Flu: We use age-specific excess respiratory death rates associated with the Spanish Flu pandemic in Kentucky 1918-1919~\cite{Viboud2013} as an approximation for the risk factors $g(0, A)$ for the Spanish Flu.
    To obtain the other risk factors $g(V{>}0, A)$, we assume a constant vaccine efficacy for all age groups.
    \item Flat Risk: Simulates of a disease where all age groups have the same severe-case risk.
\end{inparaenum}
\par
\looseness=-1 The other factors of the severity factorisation~\eqref{eqn:severity_factorisation}, $f^0$, $f^1_{\tilde \pi}$ and $h^V$, are assumed to be the same as for COVID-19 (shown in Figure~\ref{fig:severity_factorisation}).
The risk profiles are normalised such that the cumulative severe-case incidence is equal under the \texttt{Uniform} vaccine allocation strategy.
\par 
Besides the vaccine allocation strategies shown in Section~\ref{sub:counterfactual_vaccine_allocation_strategies} we consider two additional strategies that take the altered risk profiles into account:
\begin{inparaenum}[(i)]
    \item \texttt{RiskRanked}: We prioritise age groups in descending order of the risk factors.
    \item \texttt{RiskRankedReversed}: We prioritise age groups in ascending order of the risk factors.
\end{inparaenum}
\par
The factual vaccine uptake rate is influenced by the age-specific risk structure of COVID-19.
Since the elderly have a higher severe-case risk, they have more incentive to get vaccinated.
To remove this bias from the simulation set-up, we assume a flat vaccine uptake rate of $90\,\%$ willingness to receive all three vaccine doses across all age groups for the Flat and Spanish Flu risk profiles.
\par
\begin{figure}[htb]
    \centerline{
        \includegraphics[width=\singleColWidth]{figs/risk_profile_exp_overview4.pdf}
    }
    \caption{
        \looseness=-1
        \textbf{Comparison of risk profiles or disease types.}
        The top row (a-c) shows the age-specific risk factors for the unvaccinated $g(V{=}0, A)$ for three considered disease types.
        The vaccine efficacy is assumed to be independent of age and the same across all disease types.
        The risk factors are normalised across disease types to lead to the same number of severe cases under the \texttt{Uniform} vaccine allocation strategy.
        The other plots show the cumulative incidences of infections (c) and severe cases (d) under different vaccine allocation strategies between \DTMdisplaydate{2020}{12}{20}{-1} and \DTMdisplaydate{2021}{04}{11}{-1} (third wave).
        The whiskers show the $95\,\%$ credible intervals.
    }
    \label{fig:risk_profile_exp}
\end{figure}
There are no differences between diseases regarding the cumulative \emph{infection} incidence under the four initial vaccine allocation strategies, \textit{i.e.}\ \texttt{Factual},  \texttt{Uniform}, \texttt{ElderlyFirst} and \texttt{YoungFirst}.
Of those, the strategies that prioritise young and middle age groups are most effective at preventing infection spread during the third wave, as discussed in Section~\ref{sub:counterfactual_vaccine_allocation_strategies}.
For COVID-19, the \texttt{RiskRanked} and \texttt{RiskRankedReversed} strategies are most similar to the \texttt{Factual} and \texttt{YoungFirst} strategies, respectively, and lead to similar infection incidences.
For the Flat Risk, \texttt{RiskRanked} and \texttt{RiskRankedReversed} are identical to the \texttt{Uniform} strategy and lead to the same infection outcome.
For the Spanish Flu, the most effective strategy for preventing infection spread is \texttt{RiskRanked} as it prioritises age groups 20-29 and 30-39 which tend to have the highest base reproduction numbers during the third wave (see Figure~\ref{fig:reproduction_number});
Conversely, \texttt{RiskRankedReversed} is the least effective.
\par %
For COVID-19, the severe-case incidence is lowest for strategies that prioritise high-risk age groups and lowest for strategies that do the reverse, as discussed in Section~\ref{sub:counterfactual_vaccine_allocation_strategies}.
The same can be said for the Spanish Flu, however, the adversarial strategy \texttt{RiskRankedReversed} leads to even worse outcomes.
For COVID-19, younger and middle age groups have high base reproduction numbers during the third wave and the elderly have high severe-case risk.
For the Spanish Flu, however, the age groups with high severe-case risk and high base reproduction numbers are the same (20-29 and 30-39).
Therefore, the \texttt{RiskRankedReversed} strategy is adversarial in two ways: it both leads to a high infection and severe-case incidence.
\par
\looseness=-1 In summary, this simulation shows that, even when taking infection dynamics into account, following a strategy in which people most at risk are vaccinated first leads to the least amount of severe cases. 
However, we remark that the difference in severe cases is not as large as one could assume based on the difference in infection-fatality ratio alone. 
For COVID-19, a difference of $20$ years increases the fatality by a factor of~$10$, but different vaccination strategies only differ at maximum by a factor of~$4$. 
The protection against infection granted by the vaccines and the subsequent contribution to the mitigation of the epidemic waves decreases the differences of the outcomes of the different strategies. 

\subsection{Impact of immunity waning} %
\label{sub:impact_of_immunity_waning}
\label{sub:impact_of_waning_time}
We investigate the influence of immunity waning on the infection- and severe-case-incidence.
We compare three settings for the timescale at which immunity weakens:
\begin{inparaenum}[(i)]
    \item Regular: We use the waning function derived from the results reported in \cite{tartof2021}.
    \item No Waning: We assume the vaccine efficacy against infection stays constant at the maximum.
    \item Fast: The waning is $25\,\%$ faster than regular, \textit{i.e.} it takes $25\,\%$ less time until the vaccine efficacy is halved.
\end{inparaenum}
\par
The other factors of the severity factorisation~\eqref{eqn:severity_factorisation}, $f^0$, $f^1_{\tilde \pi}$ and $g$, are assumed to be the same as before (see~Figure~\ref{fig:severity_factorisation}).
For each setting of the waning function we run the four vaccine allocation strategies discussed in Section~\ref{sub:counterfactual_vaccine_allocation_strategies}.
\par %
\begin{figure}[htb]
    \centerline{
        \includegraphics[width=\singleColWidth]{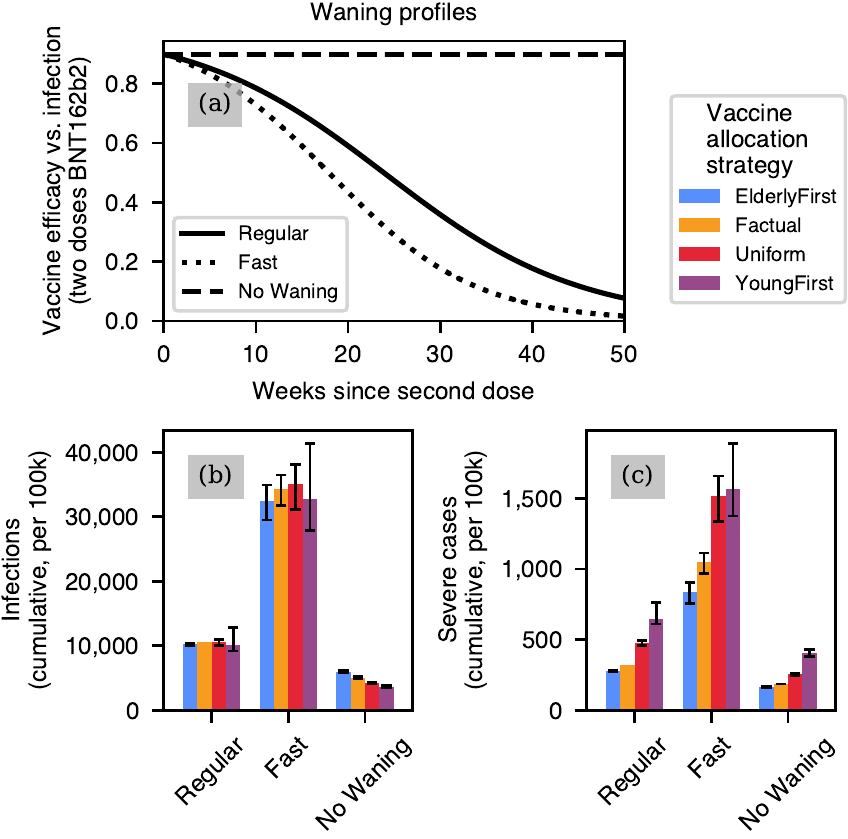}
    }
    \caption{
        \textbf{Comparison of waning timescales.}
        Plot (a) shows the considered immunity waning timescales.
        In the fast waning scenario, we assume it takes $25\,\%$ less time until vaccine efficacy against infection is halved compared to the timescale reported in~\cite{tartof2021}.
        The bottom row shows the cumulative infection (b) and severe-case (c) incidences under different vaccine allocation strategies for each waning timescale.
        We sum all cases from \DTMdisplaydate{2020}{12}{20}{-1} to \DTMdisplaydate{2021}{12}{25}{-1}.
        The whiskers show the $95\,\%$ credible intervals.
    }
    \label{fig:waning_exp}
\end{figure}%
The cumulative infection incidence increases as the speed of waning increases, as shown in Figure~\ref{fig:waning_exp} (b).
This difference is mainly driven by the behaviour of the fourth wave, where the average times since the last dose are highest.
The different strategies perform similarly: the differences in effectiveness of preventing infection spread between strategies is small compared to the differences between waning profiles.
The severe-case incidence follows the same pattern: the faster the waning, the higher the severe-case incidence (Figure~\ref{fig:waning_exp} (c)). 
\par 
These results highlight the influence of immunity waning on the infection dynamics. 
A relatively moderate reduction by $25\,\%$ of the waning timescale leads to an increase in infection- and severe-case-incidences by a factor 3 over the time span of a year.
This illustrates two points: firstly, for accurate infection dynamics predictions, it is crucial to have a good estimate of the waning curve.
Secondly, it shows the importance of regular vaccinations, counteracting the waning effect.

\section{Discussion} %
\label{sec:discussion}
\noindent We have built a model of how the severe-case probability depends on relevant factors such as age and vaccination status.
This model uses data, simulation and prior knowledge in a modular fashion and combines both parameter inference and prediction.
We used data on infections and severe cases to retrospectively evaluate different strategies in a realistic setting and explore counterfactual scenarios.
We were able to simulate the impact of increasing vaccine uptake by age group. %
The modular structure of our approach also allowed us to evaluate the behaviour of different types of diseases and the role of immunity waning.
\par 
\looseness=-1 Previous work that retrospectively evaluated vaccination campaigns focused mainly on estimating the overall success in reducing infections, hospitalisations, and deaths by modelling counterfactual scenarios with fewer or no vaccinations~\cite{vilches_estimating_2022, haas_infections_2022}.
In addition, there have been simulation studies conducted before the start of the vaccination campaign that, similar to our work, evaluate different age-dependent vaccine allocation strategies~\cite{bubar_model-informed_2021, matrajt_vaccine_2021, foy_comparing_2021}.
The forward-looking studies find that in general, the middle age groups have the biggest impact on \emph{infection} incidence (\textit{e.g.}\ Figure 1 in~\cite{bubar_model-informed_2021}, for demographics corresponding to the United States), which is in agreement with our findings (see Figure~\ref{fig:acceptance_exp_overview} (left)).
However, in these studies, such differences in contribution to spread stem from assumptions on the contact matrix~\cite{bubar_model-informed_2021, foy_comparing_2021} or prior knowledge of susceptibility~\cite{bubar_model-informed_2021, matrajt_vaccine_2021, foy_comparing_2021}, whereas in our method, we estimate the age-dependent reproduction numbers directly from infection data.
The agreement suggests that both approaches lead to qualitatively similar results.
Other studies investigating age-dependent transmission inferred from observed cases~\cite{davies_age-dependent_2020} or seropositivity data~\cite{stringhini_seroprevalence_2020} also find that middle age groups have the largest impact on transmission.
In the present work, by retrospectively taking observed infection data into account, we were additionally able to show that the timing of the vaccination campaign relative to non-pharmaceutical interventions is crucial.
In the third wave for instance, with partially open schools in Israel, vaccinating the young would have been most effective at preventing infections; whereas, in the fourth wave, with school holidays during the start of the wave, this strategy would have been among the least effective (see Figure~\ref{fig:policy_exp_cumulative}).
Nevertheless, when it comes to minimising \emph{severe cases}, all studies agree that (under realistic parameter settings~\cite{matrajt_vaccine_2021}) prioritising the elderly is most effective due to the large difference in the infection-fatality ratio~\cite{bubar_model-informed_2021, matrajt_vaccine_2021, foy_comparing_2021} (see Figures~\ref{fig:policy_exp_cumulative} and \ref{fig:acceptance_exp_overview} (right)). 
\par 
While we were able to include many factors relevant to severe cases and infections, our method has some limitations.
Our approach relies on assumptions, which are only approximately correct and difficult to test in practice.
We assume \emph{causal sufficiency}~\cite{Pearl2009} for the variables in our causal model, which rules out confounding between the variables we intervene on (vaccination times) and the outcome (severe cases).
However, we do expect some confounding in practice: at-risk groups like healthcare workers or patients with pre-existing conditions may have a higher incentive to get vaccinated~\cite{Mulberry2021}.
This could break the assumption of homogeneous subgroups based on age, vaccination status and waning time and lead to overestimated risk factors for the vaccinated $g(V{>}0, A)$.
\par
In our study, we quantify the number of infections and severe cases, where we weigh all cases equally across age groups.
However, to quantify the impact on society more accurately we could consider more factors such as differences in predisposition to long COVID~\cite{crook2021long}.
Differences in length and intensity of suffering could also be captured by metrics such as quality-adjusted life years~\cite{nord1999cost}.
However, such metrics are difficult to estimate in practice.
\par 
In the present work, we do not model behavioural or policy responses in our counterfactual scenarios explicitly.
However, high infection incidences increase perceived risk in the population and prompt voluntary health-protective behaviour~\cite{Doenges2022}.
Governments also react to changes in infection incidence by introducing or relaxing non-pharmaceutical interventions.
Both effects tend to reduce (resp.\ increase) the effective reproduction number during high (resp.\ low) incidence periods.
Therefore, we expect our infection- and severe-case-incidences to be overestimated for high-incidence periods and underestimated for low-incidence periods.
Our results should thus be interpreted as counterfactual vaccine allocation scenarios while keeping behavioural and government responses fixed.
\par 
We are also neglecting possible selection biases through differences in testing frequencies between age groups.
School children may be tested more frequently than other age groups during school weeks, which could lead to overestimating their contribution to infection spread and consequently the effect of vaccinating children.
\par
\looseness=-1 Besides approaching the problem of evaluating counterfactual vaccine allocation strategies, this study illustrates a more general problem in causal inference.
Causal models typically require joint observations of all relevant variables in order to evaluate counterfactual statements.
However, in realistic settings, data is often a limiting factor.
In the present study, two crucial parts were not observed: waning times of the severely ill and latent factors related to infection dynamics (such as the base reproduction numbers).
By assuming a factorisation for the severity mechanism~\eqref{eqn:severity_factorisation} and incorporating an SEIR-like model together with literature estimates for waning, we were able to address these limitations.
In our case, we leveraged domain expertise to merge the different sources of knowledge---namely data, simulation and literature estimates---into one model.
However, while there were some first steps in this direction~\cite{mejia2022obtaining, gresele2022causal}, how to merge information from different sources or datasets into a single causal model without strong assumptions stemming from domain knowledge is still an open question.
\par %
\looseness=-1 Through our simulation-assisted causal model, we show how interactions between different elements of a pandemic such as vaccinations, immunity waning, age-dependent infection spread and risk can be effectively captured.
While we have applied our method retrospectively, with parameter inference on observational data, this does not limit its applicability.
Each of the sub-modules can be replaced with appropriate assumptions on parts of the system which are yet unknown, as shown in Sections~\ref{sub:simulating_other_disease_types} and \ref{sub:impact_of_immunity_waning}.
For example, when COVID-19 vaccines were approved initially, it was not yet clear how long immunity against infection or severe course would last.
Thus, besides informing the rollout of COVID-19 vaccination campaigns, we hope that our method can help in future pandemics where the relevant factors may not yet be jointly measured or known from the literature.
\section{Materials and Methods} %
\label{sec:methods}

\subsection{Target function} %
\label{sub:target_function}
The target function $s(\tilde \pi = \tilde P(T_1, T_2, T_3|A))$ describes the relationship between the counterfactual vaccination policy $\tilde \pi$ and the resulting expected number of severe cases.
It can be written as
\begin{linenomath*}  %
    \begingroup
    \allowdisplaybreaks
    \begin{align}
    	&\ s(\tilde \pi = \tilde P(T_1, T_2, T_3|A)) \nonumber \\
    	&= M  D\, \EE[S|\ddo(T_1, T_2, T_3 \sim \tilde P(T_1, T_2, T_3|A) )] \\
    	&=
    	D
    	\sum_a P(a) 
        \sum_{t=1}^M f^0(t) f^1_{\tilde \pi}(a, t) \nonumber\\*
    	& \quad \times  \left[
    	\sum_{t_1=t+1}^{M+1} \sum_{t_2=t+1}^{M+1} \sum_{t_3=t+1}^{M+1} \tilde P(t_1, t_2, t_3|a) \,
    	\, g(0, a) \right. \nonumber\\*
    	 & \quad +
    	\sum_{t_1=1}^{t} \sum_{t_2=t+1}^{M+1} \sum_{t_3=t+1}^{M+1} \tilde P(t_1, t_2, t_3|a) \,
    	\, g(1, a) \, h^1(t-t_1)\nonumber\\*
    	 & \quad +
    	\sum_{t_1=1}^{t} \sum_{t_2=1}^{t} \sum_{t_3=t+1}^{M+1} \tilde P(t_1, t_2, t_3|a) \,
    	\, g(2, a) \, h^2(t-t_2)\nonumber\\*
    	 & \quad \left. +
    	\sum_{t_1=1}^{t} \sum_{t_2=1}^{t} \sum_{t_3=1}^{t} \tilde P(t_1, t_2, t_3|a) \,
    	\, g(3, a) \, h^3(t-t_3) \right]
    	\label{eqn:target_function}
    \end{align}
    \endgroup
\end{linenomath*}
where $D$ is the total population.
For notational convenience, we treat individuals who have not received a certain dose by setting the respective time of vaccination to $t_i = M+1$.
The full derivation is shown in Supplement~\ref{app:target_function}.

\subsection{Estimating the severity mechanism factors} %
\label{sub:estimating_the_severity_conditional_factors}

\paragraph{Risk factors for the unvaccinated} %
\label{par:estimating_risk_factors_for_the_unvaccinated}
Estimating the factorisation \eqref{eqn:severity_factorisation} is ill-posed because the overall scale of the factors $g$, $h^V$ and $f^0$ is not well-defined.
We can double $f^0$ and halve $g$ and end up with the same value for the severity mechanism. 
We remove this ambiguity by setting 
\begin{equation}
	g(0, a^*) = 1
\end{equation}
for some age group $a^*$.
By common convention, we choose the age group $60\text{-}69$ as the reference group such that all other risk factors are relative to $g(0, a^*)$.
Note that there is no waning for $V=0$, hence $P(S{=}1|V{=}0, A, T, W) = P(S{=}1|V{=}0, A, T)$.
We can then estimate the other risk factors of the unvaccinated by
\begin{equation}
	\hat g(0, a) = \frac{\EE_T [P(S{=}1|V{=}0, A{=}a, T)]}{\EE_T [P(S{=}1|V{=}0, A{=}a^*, T)]}.
\end{equation}

\paragraph{Immunity waning curve} %
\label{par:immunity_waning_curve}
\looseness=-1 The vaccine efficacy against infection as a function of time since the administration of the second dose of the BioNTech vaccine is reported in~\cite{tartof2021} for discrete time periods up to 6 months. 
To this data, we fit a logistic curve that tends towards zero efficacy as time increases.
For the waning curves after 1 and 3 doses, we assume the same functional relationship as for the second dose, but scale the function such that under full protection the efficacy is 75\% and 95\%, respectively (efficacy under full protection is around 90\% after two doses).
\par
We can use this to derive the waning function $h^V(W)$.
First observe that the severe-case probability can be separated into two processes: 
\begin{linenomath*}  %
    \begin{align}
    	P(S{=}1|V, A, T, W) &= 
    	\underbrace{P(I{=}1|V, A, T, W)}_{\text{(a)}}
    	\ 
    	\underbrace{P(S{=}1|V, A, I{=}1) }_{\text{(b)}}
    	\label{eqn:severity_processes}
    \end{align}
\end{linenomath*}
\begin{inparaenum}[(a)]
	\item the probability of being infected and \label{infection_process}
	\item the probability of developing a severe case once infected. \label{severity_process}
\end{inparaenum}
We assume that the probability of having a severe case once infected only depends on $V$ and $A$ and the immunity against severe courses does not significantly wane over time, as reported in~\cite{tartof2021}.
Now, let $\mathrm{VE}^v(w)$ be the vaccine efficacy $w$ weeks after receiving the $v^\mathrm{th}$ dose:
\begin{equation}
	\mathrm{VE}^v(w) = 1 - \frac{P(I{=}1|V{=}v, A{=}a, T, W{=}w)}{P(I{=}1|V{=}0, A{=}a, T, W{=}w)}
\end{equation}
where we assume that the efficacy against infection is the same for all age groups and constant over time $T$.
Then note, using equation~\eqref{eqn:severity_processes},
\begin{linenomath*}  %
    \begin{align}
    	&\, \frac{P(S{=}1|V{=}v, a, t, w)}{P(S{=}1|V{=}0, a, t, w)} \nonumber \\
    	= &\, \frac{P(S{=}1|V{=}v, a, I{=}1)}{P(S{=}1|V{=}0, a, I{=}1)}
    	\underbrace{\frac{P(I{=}1|V{=}v, a, t, w)}{P(I{=}1|V{=}0, a, t, w)}}_{1 - \mathrm{VE}^v(w)} \\
    	= &\, \frac{f^0(t) g(v,a) h^v(w)}{f^0(t) g(0,a)}
    	= \frac{g(v,a)}{g(0,a)} h^v(w) \\ 
    	\Rightarrow &\, h^v(w) = \frac{g(0,a)}{g(v,a)}
    	\frac{P(S{=}1|V{=}v, a, I{=}1)}{P(S{=}1|V{=}0, a, I{=}1)}
    	(1 - \mathrm{VE}^v(w)) . 
    \end{align}
\end{linenomath*}
Note that since we are considering the factual vaccine allocation strategy $\pi$ the correction factor is $f^1_{\pi}(A, T) = 1$.
By definition we have $h^v(0) = 1$ and combining the above expression for $W=w\ge0$ and $W=0$ we get
\begin{equation}
	\frac{h^v(w)}{h^v(0)} = \frac{1 - \mathrm{VE}^v(w)}{1 - \mathrm{VE}^v(0)}
	\Rightarrow
	h^v(w) = \frac{1 - \mathrm{VE}^v(w)}{1 - \mathrm{VE}^v(0)} .
	\label{eqn:infection_waning_function}
\end{equation}

\sloppy
\paragraph{Risk factors for the vaccinated} %
\label{par:estimating_risk_factors_for_the_vaccinated}
In our data~\cite{Israel2021} we can only observe $P(S{=}1|V,A,T)$ since we do not have data on severe outcomes as a function of the time since the last dose $W$.
However, we do have data on the distribution of times since the last dose was received $P(W|V, A, T)$.
This allows us to use the waning function \eqref{eqn:infection_waning_function} to estimate the risk factors for $V{=}v>0$ since
\begin{linenomath*}  %
    \begin{align}
    	& P(S{=}1|V{=}v, a, t)
    	= \sum_w P(S{=}1|V{=}v, a, t, w) P(w|V{=}v, a, t) \\ 
    	&=  f^0(t) g(v,a) \sum_w h^v(w) P(w|V{=}v, a, t).
    \end{align}
\end{linenomath*}
This motivates the following estimator:
\begin{equation}
	\hat g(v, a) 
	= \frac{
	\EE_T \left[ \frac{P(S{=}1|V{=}v, A{=}a, T)}{\EE_{W|V=v, A=a, T}[h^v(W)] } \right]
	}{
	\EE_T [P(S{=}1|V{=}0, A{=}a, T)]
	}
	 \, \hat g(0, a) ,
\end{equation}
\textit{i.e.}, we correct for the waning that occurred in the population to estimate the risk factor under full immunity.
\fussy

\paragraph{Time dependence} %
\label{par:estimating_the_time_dependence}
After correcting for immunity waning and differences in risk factors, we can estimate the overall time dependence:
\begin{equation}
	\hat f(t) = \EE_{V, A|T=t} \left[ 
	    \frac{P(S{=}1|V, A, T{=}t)}{\hat g(V, A) \EE_{W|V, A, T{=}t}[h^V(W)]}
	\right].
\end{equation}

\paragraph{Infection dynamics correction factor} %
\label{par:estimating_the_infection_dynamics_correction_factor}
For the estimation of $f^1_{\tilde  \pi}(A,T)$, we consider the following: let $P_\pi(S{=}1|V, A, T, W)$ be the severity mechanism under the observed vaccine allocation strategy $\pi$ and $P_{\tilde \pi}(S{=}1|V, A, T, W)$ under the post-intervention vaccine allocation strategy $\tilde \pi$.
Then, using equation~\eqref{eqn:severity_processes},
\begin{align}
	\frac{P_{\tilde \pi}(S{=}1|V, A, T, W)}{P_{\pi}(S{=}1|V, A, T, W)}
	=&\, \frac{P_{\tilde \pi}(I{=}1|V, A, T, W)\ P(S{=}1|V, A, I{=}1)}{P_{\pi}(I{=}1|V, A, T, W)\ P(S{=}1|V, A, I{=}1)} \nonumber \\ 
	=&\, \frac{P_{\tilde \pi}(I{=}1|V, A, T, W)}{P_{\pi}(I{=}1|V, A, T, W)} \label{eqn:correction_factor_derivation}
\end{align}
where we have used that the process of going from infected to severely ill (\ref{severity_process}) does not depend on the vaccine allocation strategy.
From the factorisation of the severity mechanism~\eqref{eqn:severity_factorisation} and using $f^1_{\pi}(A, T){=}1$ it follows that
\begin{equation}
    \frac{P_{\tilde \pi}(S{=}1|V, A, T, W)}{P_{\pi}(S{=}1|V, A, T, W)} = f^1_{\tilde \pi}(A, T) .
\end{equation}
The assumption that $f^1_{\tilde \pi}$ only depends on $A$ and $T$ means we assume that the change in infection probability due to the infection dynamics is independent of $V$ and $W$.
Hence, the correction factor is the relative change in weekly infection probability for each age group under the counterfactual vaccine allocation strategy:
\begin{equation}
   f^1_{\tilde  \pi}(A,T)  = \frac{P_{\tilde \pi}(I{=}1|V, A, T, W)}{P_{\pi}(I{=}1|V, A, T, W)}.
\end{equation}

\subsection{Modelling infection dynamics} %
\label{sub:modelling_infection_dynamics}

\sloppy To estimate the effect of changing the vaccine allocation strategy on the infection dynamics we first infer the parameters of a Bayesian SEIR-like model to describe infections $P_{\pi}(I{=}1|V, A, T, W)$ under the observed policy $\pi$.
We then rerun the model with the inferred reproduction numbers under the counterfactual strategy $\tilde \pi$ to obtain an estimate of $P_{\tilde \pi}(I{=}1|V, A, T, W)$.
The correction factor $f^1_{\tilde \pi}(A, T)$ is given by the ratio of these two infection probabilities (equation~\eqref{eqn:correction_factor}).
\begin{figure}[h!]
    \centering
    \includegraphics[width=\singleColWidth]{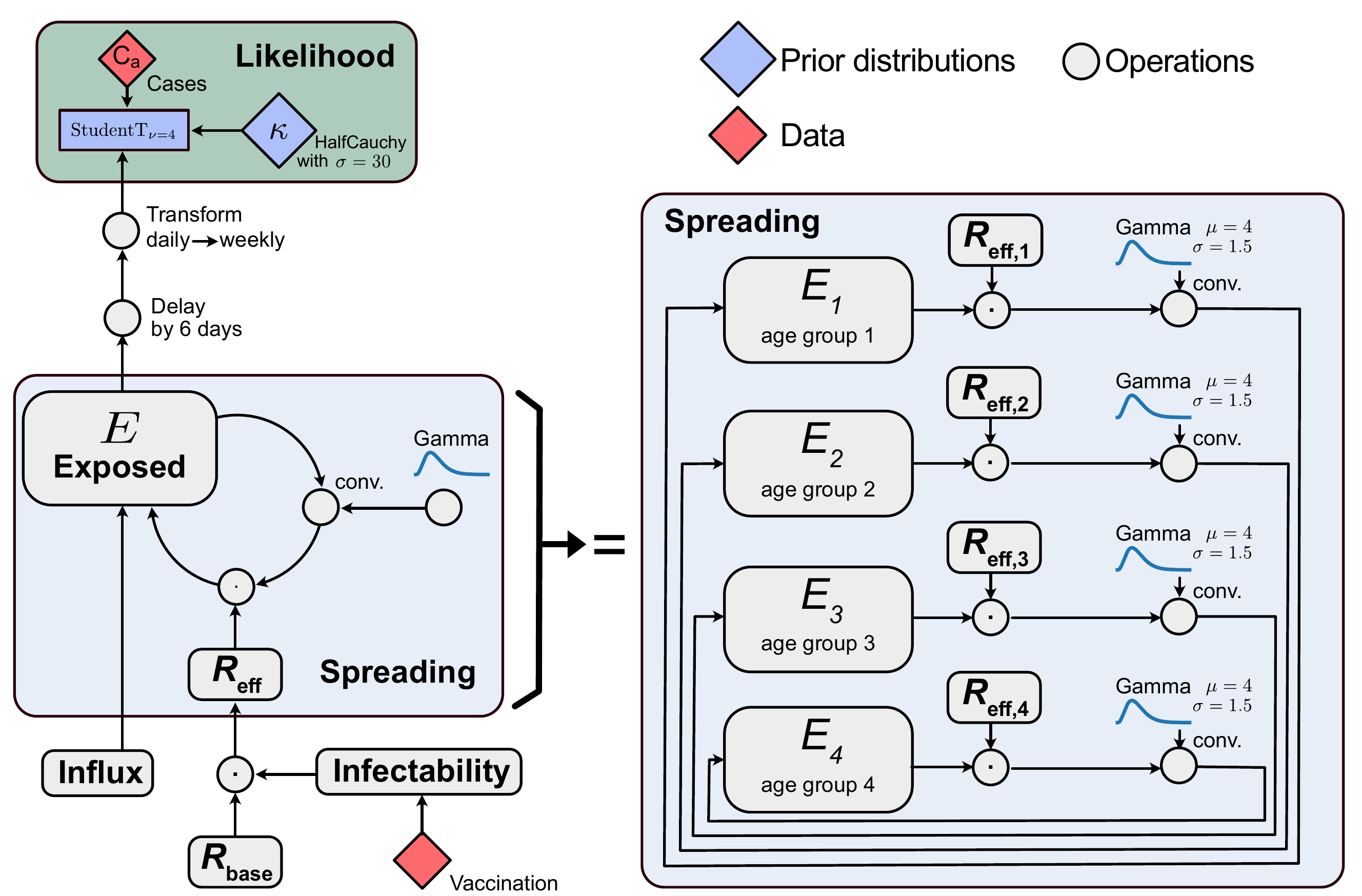}
    \caption{
    \textbf{Overview of the infection dynamics model.} 
    The dynamics model infers the effective $R_\text{eff}$ and base reproduction number as a function of time given the factual vaccine allocation strategy and the number of observed cases $C_a$.
    After inference, the model is used to predict the number of cases under a counterfactual strategy. 
    The right side corresponds to equation~\eqref{eq:spread}.
    Only 4 out of the 9 age groups are shown here.
    }
    \label{fig:model_overview_dynamics}
\end{figure}

\subsubsection{The SEIR-like dynamics}
In our model each age group $a$ has its own compartment, and the dynamics follows a discrete renewal process \cite{fraser_estimating_2007, brauner_inferring_2020} determined by the \emph{effective reproduction number} $R_{\text{eff}, a}(t_\text{day})$.
These dynamics are discretized with a one-day step, to allow for enough resolution to model the generation interval. 
The data is only available on a weekly basis, which will require later to sum the cases over one week.
We fit one reproduction number per age group.
The infections from one age group to another are encoded by a contact matrix $\mathbf{C}$.
The latent period is modelled by a kernel $g(\tau)$ which is normalized to 1: 
\begin{linenomath*}  %
    \begin{align}
        E_{a}(t_\text{day}) &=  \sum_{a'=1}^9 \sqrt{R_{\text{eff}, a}(t_\text{day})} C_{a,a'}  \sqrt{R_{\text{eff}, a'}(t_\text{day})} \nonumber \\
        &\, \quad \sum_{\tau=0}^{10} E_{a'}(t_\text{day}-1-\tau) g(\tau), \label{eq:spread} + h_a(t_\text{day}),\\
        S_{a}(t_\text{day}) &= S_{a}(t_\text{day}-1) - E_{a}(t_\text{day}-1),\\
        g(\tau) &= \text{Gamma}(\tau;\mu{=}4,\sigma{=}1.5).
    \end{align}
\end{linenomath*}
$E_{a}(t_\text{day})$ is the number of newly exposed people on day $t_\text{day}$, who will later become infectious;
It is implicitly modelled by the generation interval kernel $g$. We assume here a mean generation interval of 4 days \cite{pung_serial_2021, hart_generation_2022}.
$S_{a}(t_\text{day})$ is the number of susceptible people and $h_a(t_\text{day})$ is an external influx (see below). 
\par
We assume that changes in the reproduction number symmetrically affect both the infectiousness and infectiability.
This is achieved by multiplying the square-root of the reproduction number to the contact matrix from both sides in equation~\eqref{eq:spread}. 

\subsubsection{The contact matrix}

The contact matrix models the amount of contacts between age groups.
There are two extreme cases: 
\begin{enumerate}
    \item There are no contacts between age groups.
    In this case, the contact matrix would simply be the identitiy matrix: $\mathbf{C} = \mathbbm{1}$. 
    \item The contact between different age groups is the same as within age groups: all-to-all connectivity.
    Let $D^{(A=a)}$ be the population of age group $a$ and $D = \sum_a D^{(A=a)}$ be the total population.
    Then, in this case, contact matrix would be $\mathbf{C} = \vec{\rho} \cdot \vec{\text{1}}^T$, where $\vec{1}^T = \left(1, 1, \dots, 1\right)$ and $\vec{\rho}^T = \left(\rho_1, \rho_2, \dots \right) = \left( \frac{D^{(A=1)}}{D}, \frac{D^{(A=2)}}{D}, \dots\right)$ is the vector of the population share in each age group.  
\end{enumerate}
The reality lies somewhere between these two cases.
A reasonable interpolation between these extremes should ensure that the largest eigenvector of the contact matrix stays $\vec{\rho}$ and that the largest eigenvalue is 1.
This ensures that under a reproduction number of 1 the case numbers are constant and proportional to the relative population $\vec{\rho}$ of each age group.
These requirements are met by the following matrix: 
\begin{linenomath*}  %
    \begin{align}
        \mathbf{C} &=  \left(1-\gamma\right) \mathbbm{1} + \gamma \vec{\rho} \cdot \vec{\text{1}}^T \nonumber\\
        &=\begin{pmatrix}
            \gamma\rho_1+\left(1-\gamma\right)  & \gamma \rho_1  & \gamma\rho_1  & \dots\\
             \gamma \rho_2  &  \gamma\rho_2+\left(1-\gamma\right) & \gamma\rho_2 & \dots\\
             \gamma \rho_3 & \gamma \rho_3 &\gamma\rho_3+\left(1-\gamma\right) & \dots \\
             \vdots & \vdots & \vdots &\ddots
        \end{pmatrix} .
    \end{align}
\end{linenomath*}
The parameter $\gamma$ determines the contact mixing between age group: at 1 we have all-to-all connectivity, at 0 there is no contact between age groups.
 
\subsubsection{The external influx}

To account for some infections occurring due to infected travellers entering Israel, we add a random number of infections distributed over each week:
\begin{linenomath*}  %
    \begin{align}
        h_a(t_\text{day}) &= h^*_a(t = k_\text{week}(t_\text{day}) )/7,\\
        h^*_a(t) &\sim \text{Weibull}\left(\lambda = 0.1\cdot \frac{\text{pop}_a}{10^6},  k = 0.3\right) \quad \forall a, \forall t,
    \end{align}
\end{linenomath*}
where $t$ is indexing the weeks included in our analysis, $k_\text{week}(t_\text{day})$ is mapping a day $t_\text{day}$ to the corresponding week $t$.
We chose a Weibull distribution because the long tails allow the occurrence of mass spreading events.
The parameters of the Weibull distribution are chosen such that on average 0.1 infections per million inhabitants per day occur from external influx.
This is about a fifth of the lowest incidence during the analysis period.
The median of the distribution is only at 0.003 infections per million inhabitants per day because of the long tails of the distribution.

\subsubsection{The effective reproduction number}
The effective reproduction number $R_{\text{eff}, a}(t_\text{day})$  depends on (1) the base reproduction number $R_{\text{base}, a}(t_\text{day})$, which encodes the amount of social distancing at time $t_\text{day}$ and (2) the infectiability term $\text{Infectability}_a (t_\text{day})$ which encodes the acquired immunity of the susceptible population of age group $a$
\begin{linenomath*}  %
    \begin{align}
        R_{\text{eff}, a}(t_\text{day}) =  R_{\text{base}, a}(t_\text{day}) \cdot \text{Infectability}_a (t_\text{day}) .
    \end{align}
\end{linenomath*}

\subsubsection{The infectability}
\looseness=-1 The $\text{Infectability}_a$ is the fraction of reduced spread due to acquired immunity in age group $a$.
It is 1 in a completely non-immune population and reduces with vaccination.
Concretely, it is modelled as:
\begin{linenomath*}  %
    \begin{align}
        \text{Infectability}_a(t_\text{day}) =&\,  \text{Unv}_a(t) \nonumber \\
         &+ \text{Vacc}^1_a(t) \left(1- \mu^1_a \text{W}^1_{\text{eff}, a} \left(t\right)\right) & \nonumber\\  
         &+ \text{Vacc}^2_a(t) \left(1- \mu^2_{a} \text{W}^2_{\text{eff}, a} \left(t\right)\right) & \nonumber\\ 
         &+ \text{Vacc}^3_a(t)\left(1- \mu^3_a \text{W}^3_{\text{eff}, a} \left(t\right)\right)\\
         & \text{ with }  t = k_\text{week}(t_\text{day}), 
    \end{align}
\end{linenomath*}
where $\text{Unv}_a(t)$, $\text{Vacc}^1_a(t)$, $\text{Vacc}^2_a(t)$ and $\text{Vacc}^3_a(t)$ are the fractions of unvaccinated, once-, twice- and three-times-vaccinated, respectively, for each age group.
Here $t=k_\text{week}(t_\text{day})$ is the week corresponding to the day $t_\text{day}$.
$\mu$ denotes the corresponding protection against infection ($0 < \mu < 1$, $\mu = 0$ corresponds to no protection).
We assume that the protection is $70\,\%$, $90\,\%$ and $95\,\%$ directly after the first, second and third dose, respectively~\cite{tartof2021}.
These vaccination fractions are modelled in a weekly manner because the data is only available on a weekly basis. 
$\text{W}^v_{\text{eff}, a}(t)$  denotes the effective group-wide waning of the immunity for the group that has been vaccinated $v$ times. 
It is 1 if the whole age group had been vaccinated a few days ago and decreases with time. 
\par
The effective waning of the group-wide immunity at time $t$ is calculated 
by building an average over all individuals who have received $v$ doses before time $t$ weighted by their individual waning factor:
\begin{linenomath*}  %
    \begin{flalign}
        \text{W}^v_{\text{eff}, a} (t) &=  \frac{\sum_{\tau=0}^t \text{n\_Vacc}^v_{a, t}(\tau) 
    \text{VE}_\text{norm}(t-\tau)}{\sum_{\tau=0}^t\text{n\_Vacc}^v_{a, t}(\tau) }
    \end{flalign}
\end{linenomath*}
where $\text{n\_Vacc}^v_{a, t}(\tau)$ are the newly vaccinated at time $\tau$ who have received $v$ vaccinations by time $\tau$ and $\text{VE}_\text{norm}(w) = \frac{\text{VE}(w)}{\text{VE}(0)}$ is the normalized vaccine efficacy $w$ weeks after the last dose~\cite{tartof2021}.
$\text{n\_Vacc}^v_{a, t}(\tau)$ and $\text{Vacc}^v_a(t)$ are both obtained from published data from Israel~\cite{Israel2021}.

\subsubsection{The base reproduction number}
The base reproduction number $R_{\text{base}, a}(t_\text{day})$ is assumed to be a slowly changing factor as a function of time.
It is modelled as a superposition of logistic change points $\gamma(t_\text{day})$ every 21 days, which are parameterised by the transient length of the change points $l$, the date of the change point $d$ and the effect of the change point $\Delta \gamma^*$.
The subscripts $n$ denote the discrete enumeration of the change points:
\begin{linenomath*}  %
    \begingroup
    \allowdisplaybreaks
    \begin{align}
         R_{\text{base}, a}(t_\text{day}) &= R_{0, a} \exp \left( \sum_n \gamma_{n}(t_\text{day})\right)\label{eq:R_base}\\
         R_{0, a} &\sim \text{LogNormal}\left(\mu=1, \sigma=1\right) \quad \forall a \\
         \gamma_{n, a}(t_\text{day}) &= \frac{1}{1+e^{-4/l_{n, a} \cdot (t - d_{n, a})}} \cdot \Delta \gamma_{n, a}\label{eq:gamma_sigmoid_base}\\
        \Delta \gamma_{n, a} &\sim \mathcal{N}\left(\Delta \gamma_{n-1, a}, \sigma_{\Delta \gamma_a} \right) \quad \forall n, \forall a \nonumber \\*
        & \quad \ \text{with} \,\, \Delta \gamma_{0, a} =  \log R_{0, a}\\
        \sigma_{\Delta \gamma, a} &\sim \text{HalfCauchy}\left(0.5\right) \quad \forall a\\
        l_{n, a} &= \log\left(1+\exp(l^\dagger_{n, a})\right)\label{eq:length_cp_base}\\
    l^\dagger_{n, a} &\sim \mathcal{N}\left(4, 1\right) \quad \forall n, \forall a \quad \text{(unit is days)}\\
        d_{n,a} &= 10^\text{th} \text{ January 2021}  + 21 \cdot n + \Delta d_{n,a} \nonumber \\*
        & \quad \ \text{for } n = {0,\dots,9}\\
        \Delta d_{n_a}  &\sim \mathcal{N}\left(0, 3.5\right) \quad \forall n, \forall a \quad \text{(unit is days)}.
    \end{align}
    \endgroup
\end{linenomath*}

\subsubsection{The likelihood}
Next, we want to define the goodness of fit of our model to the sample data.
For that, the number of newly exposed people is delayed by 6 days and summed over one week because the case data is available on a weekly basis.
The likelihood of that is modelled by a Student-t distribution, which allows for some outliers because of its heavier tails compared to a normal distribution (green box in Figure~\ref{fig:model_overview_dynamics}).
The error of the Student-t distribution is proportional to the square root of the number of cases, which corresponds to the scaling of the errors in a Poisson or Negative Binomial distribution: 
\begin{linenomath*}  %
    \begin{align}
    \hat{C}_a(t) &= \sum_{t_\text{day}=6}^{13} E\left(k_\text{week}^{-1}\left(t\right)-t_\text{day}\right) \\
        C_a(t) &\sim \text{StudentT}_{\nu = 4}\left( 
            \mu= \hat{C}_g(t),
            \sigma = \kappa \sqrt{\hat{C}_a(t)+1}\right) \\
            \kappa &\sim \text{HalfCauchy}(\sigma=30).
    \end{align}
\end{linenomath*}
Here $C_a(t)$ is the measured number of weekly cases in the population of age $a$ as reported by the health authorities, whereas $\hat{C}_g(t)$ is the modelled number of cases in week $t$. 
$k_\text{week}^{-1}(t)$ transforms the week $t$ to the first day of that week.

\subsubsection{Sampling}

To estimate the parameters of the bayesian dynamical spreading model, in particular the time-dependent base reproduction number, we use Monte-Carlo sampling. 
In this way, we also obtain credible intervals of the parameters and not only the maximal likelihood estimate.
Specifically, the sampling was performed using PyMC3~\cite{salvatier_probabilistic_2016} with the NUTS sampler~\cite{hoffman_no-u-turn_2011}, which is a Hamiltonian Monte-Carlo sampler. 
\par
The chains were initialized randomly.
As random initialisation often leads to some chains getting stuck in local minimum, we run 8 chains for 150 initialisation steps and chose the 2 chains with the highest unnormalised posterior to continue tuning and sampling.
We then let these chains tune for additional 500 steps and draw 500 samples.
The maximum tree depth was set to 10. 

\subsection{Credible intervals} %
\label{app:credible_intervals}

\sloppy
\paragraph{Infections} For all quantities related to infections we sample from the Bayesian SEIR-like infection dynamics model to obtain samples of the posterior distribution $P_{\tilde \pi}(I{=}1|V, A, T, W)$.
For the credible interval we take 1000 joint samples of $P_{\tilde \pi}(I{=}1|V, A, T, W)$ to obtain samples of the final quantity such as the total sum of infections.

\paragraph{Severe cases} We compute samples for the correction factor~\eqref{eqn:correction_factor} by sampling from the posterior of the infection dynamics model 1000 times.
These joint samples of the correction factor are then propagated through the target function~\eqref{eqn:target_function} to obtain samples of the severe-case incidence.
\footnotesize
\section*{Data and Code Availability}

The epidemiological data that support the findings of this study are available from the Ministry of Health Israel~\cite{Israel2021}, \url{https://data.gov.il/dataset/covid-19}.
The population data used for estimating the age distribution in Israel are available from the United Nations World Population Prospects 2019~\cite{UN2019}, \url{https://population.un.org/wpp}.
The source code is available at \url{https://github.com/akekic/covid-vaccine-evaluation} and contains copies of all used data sources.
\section*{Acknowledgements}

We thank Sebastian Contreras and Simon Bauer for constructive discussions.
We thank Franz Paul Spitzner for comments on visualisations.
\textbf{Funding:}
This publication was supported by the German Ministry of Science and Education (BMBF) through the Tübingen AI Center (FKZ 01IS18039A) and by the German Research Foundation (Deutsche Forschungsgemeinschaft, DFG) through both the Cluster of Excellence “Machine Learning - New Perspectives for Science”, (EXC 2064, project number 390727645) and the project “Cognition of Interaction” (SFB 1528, project number 454648639).
\textbf{Author contributions:}
AK and JD wrote the software and carried out the analysis.
AK, LG, JvK and BS conceptualized the causal model. VP conceptualized the infection dynamics model.
All authors contributed to the development of the methodology and the manuscript draft.
\textbf{Competing interests:}
The authors declare that they have no competing interests.

\footnotesize
\printbibliography
\clearpage
\normalsize

\renewcommand{\appendixpagename}{Supplementary Material}
\renewcommand{\figurename}{Supp.\ Figure}

\begin{appendix}
    \appendixpage
    \counterwithin{figure}{section}
    \counterwithin{table}{section}
    
    \section{Generating vaccine allocation strategies} %
    \label{app:generating_vaccine_allocation_strategies}
    
    \subsection{Factual strategy}
    \label{app:factual_strategy}
    
    In the available dataset~\cite{Israel2021} the vaccination times for first, second and third doses are given for each age group, \textit{i.e.}\ 
    \begin{equation}
        P(T_i| A) \quad \text{for } i=1, 2, 3. \label{eqn:t_given_a_obs}
    \end{equation}
    However, this alone does not uniquely determine the distribution of waning times.
    To fully specify required joint distributions $P(T_1, T_2|A)$ and $P(T_3| T_2, A)$ we employ a greedy algorithm to
    \begin{enumerate}
        \item Maximise the number of vaccine recipients that receive their second dose 3 weeks after their initial dose (or as close to 3 weeks as possible, but not sooner).
        \item Have a minimum gap of 12 weeks between second and third dose (constraint).
        \item Satisfy~\eqref{eqn:t_given_a_obs} (constraint).
    \end{enumerate}

    \subsection{Uniform strategy}
    \label{app:uniform_strategy}
    
    For all counterfactual strategies we require the number of first, second and third doses to match the factual---but not in each age group. In other words,
    \begin{equation}
        \tilde P(T_i) = P(T_i) \quad \text{for } i=1, 2, 3. \label{eqn:t_obs}
    \end{equation} 
    where the tilde indicates probabilities in the counterfactual scenario. For the \texttt{Uniform} strategy, we require all age groups to have the same vaccination time distributions:
    \begin{equation}
        \tilde P(T_1, T_2|A) = \tilde P(T_1, T_2) \quad \text{and} \quad \tilde P(T_3| T_2, A) = \tilde P(T_3| T_2). \label{eqn:uniform_constraint}
    \end{equation}
    In summary, the greedy algorithm to determine $\tilde P(T_1, T_2|A)$ and $\tilde P(T_3| T_2, A)$ is adapted as follows:
    \begin{enumerate}
        \item Maximise the number of vaccine recipients that receive their second dose 3 weeks after their initial dose (or as close to 3 weeks as possible, but not sooner).
        \item Have a minimum gap of 12 weeks between second and third dose (constraint).
        \item Satisfy~\eqref{eqn:t_obs} (constraint).
        \item Satisfy~\eqref{eqn:uniform_constraint} (constraint).
    \end{enumerate}

    \subsection{Ranked strategies}
    \label{app:ranked_strategies}
    
    In the ranked strategies \texttt{ElderlyFirst}, \texttt{YoungFirst}, \texttt{RiskRanked} and \texttt{RiskRankedReversed} all age groups are ranked and vaccinations are assigned to the highest-ranked age group its vaccine uptake rate is reached.
    The counterfactual vaccine uptake rate per age group is the factual one for the first and second doses, \textit{i.e.}\ 
    \begin{equation}
        \sum_{t_i=1}^M \tilde P(t_i|a) \le \sum_{t_i=1}^M P(t_i|a) \quad \text{for } i=1, 2 \text{ and } \forall a. \label{eqn:VAR_12}
    \end{equation}
    The vaccine uptake rate for booster shots is slightly relaxed by 2.5\% to meet other consistency constraints:
    \begin{equation}
        \sum_{t_3=1}^M \tilde P(t_3|a) \le 0.025 + \sum_{t_3=1}^M P(t_3|a) \quad \forall a. \label{eqn:VAR_3}
    \end{equation}
    The greedy algorithm is adapted to
    \begin{enumerate}
        \item Prioritise age groups according to ranking.
        \item Maximise the number of vaccine recipients that receive their second dose 3 weeks after their initial dose (or as close to 3 weeks as possible, but not sooner).
        \item Have a minimum gap of 12 weeks between second and third dose (constraint).
        \item Satisfy~\eqref{eqn:t_obs} (constraint).
        \item Satisfy~\eqref{eqn:VAR_12} and~\eqref{eqn:VAR_3} (constraint).
    \end{enumerate}

    \section{Target function} %
    \label{app:target_function}
    
    Our goal is to compute the expected number of severe cases after intervening on the distribution of vaccination times:
    \begin{linenomath*}  %
        \begin{align}
            &\ s(\tilde \pi = \tilde P(T_1, T_2, T_3|A)) \nonumber \\
            &:= M D\ \EE[S|\ddo(T_1, T_2, T_3 \sim \tilde P(T_1, T_2, T_3|A) )]
            \label{eqn:target_function_start}
        \end{align}
    \end{linenomath*}
    where $\tilde P(T_1, T_2, T_3|A)$ is the distribution of vaccination times in the counterfactual scenario. 
    Using results for stochastic policies from~\cite{Pearl2009}, repeated application of the law of total probabilities (TP) and exploiting conditional independences (CI) implied by the causal graph (see Figure~\ref{fig:causal_graph}) we can derive an expression for~\eqref{eqn:target_function_start} in terms of known conditional probabilities:
    \begin{linenomath*}  %
        \begingroup
        \allowdisplaybreaks
        \begin{align}
            &\ s(\tilde \pi = \tilde P(T_1, T_2, T_3|A)) \nonumber \\
            &:= M D\ \EE[S|\ddo(T_1, T_2, T_3 \sim \tilde P(T_1, T_2, T_3|A) )] \\
            &= M D\ P\left(S{=}1|\ddo(T_1, T_2, T_3 \sim \tilde P(T_1, T_2, T_3|A) ) \right) \\
            &\stackrel{\text{\cite[Ch.~4.2]{Pearl2009}}}{=} M D \sum_a P(a) 
            \sum_{t_1, t_2, t_3 = 1}^{M+1} \tilde P(t_1, t_2, t_3|a) \nonumber \\*
            &\qquad
            P(S{=}1|\ddo(t_1, t_2, t_3), a) \\
            &\stackrel{\text{TP}}{=} M  D \sum_a P(a) 
            \sum_{t_1, t_2, t_3 = 1}^{M+1} \tilde P(t_1, t_2, t_3|a)
            \sum_{t=1}^M \underbrace{P(t|\ddo(t_1, t_2, t_3), a)}_{\stackrel{\text{CI}}{=} P(t)} \nonumber \\*
            &\qquad 
            P(S{=}1|\ddo(t_1, t_2, t_3), a, t) \\
            &\stackrel{\text{TP}}{=} M  D \sum_a P(a)
            \sum_{t=1}^M  \underbrace{P(t)}_{= 1/M}
            \sum_{t_1, t_2, t_3 = 1}^{M+1} \tilde P(t_1, t_2, t_3|a)  \nonumber \\*
            &\qquad 
            \sum_{w=1}^M \underbrace{P(w|\ddo(t_1, t_2, t_3), a, t)}_{\stackrel{\text{CI}}{=} P(w|\ddo(t_1, t_2, t_3), t)}
            P(S{=}1|\ddo(t_1, t_2, t_3), a, t, w) \\
            &\stackrel{\text{TP}}{=} D \sum_a P(a)
            \sum_{t=1}^M 
            \sum_{t_1, t_2, t_3 = 1}^{M+1} \tilde P(t_1, t_2, t_3|a)
            \sum_{w=1}^M P(w|\ddo(t_1, t_2, t_3), t) \nonumber \\*
            &\qquad 
            \sum_{v=0}^3 \underbrace{P(v|\ddo(t_1, t_2, t_3), a, t, w)}_{\stackrel{\text{CI}}{=} P(v|\ddo(t_1, t_2, t_3), t)}  \underbrace{P(S{=}1|\ddo(t_1, t_2, t_3), v, a, t, w)}_{\stackrel{\text{CI}}{=} P(S{=}1|v, a, t, w)} \\
            &= D \sum_a P(a)
            \sum_{t=1}^M
            \sum_{t_1, t_2, t_3 = 1}^{M+1} \tilde P(t_1, t_2, t_3|a) \sum_{w=1}^M P(w|\ddo(t_1, t_2, t_3), t) \nonumber \\*
            &\qquad  
            \sum_{v=0}^3 P(v|\ddo(t_1, t_2, t_3), t)\ P(S{=}1|v, a, t, w) \\
            &= D \sum_a P(a) 
            \sum_{t=1}^M
            \sum_{t_1, t_2, t_3 = 1}^{M+1} \tilde P(t_1, t_2, t_3|a) \sum_{w=1}^M P(w|t_1, t_2, t_3, t)  \nonumber \\*
            &\qquad 
            \sum_{v=0}^3 P(v|t_1, t_2, t_3, t)\ P(S{=}1|v, a, t, w) .
        \end{align}
        \endgroup
    \end{linenomath*}
    \looseness=-1 In the last step we use that since $\{T_1, T_2, T_3, T\}$ are the parents of $V$ and $W$ we can replace the $\ddo$-operators by conditional probabilities~\cite{Pearl2009}. Also note that for notational convenience, we set the vaccination time $t_i = M+1$ for a patient who has not received the $i^\mathrm{th}$ dose during the considered time window $t\in \{ 1, \dots, M \}$.
    \par
    The waning time $w$ (number of weeks since the last dose was received) depends deterministically on the vaccination times $t_1, t_2, t_3$: 
    \begin{align}
        P(w|t_1, t_2, t_3, t)
 		= &\,
 		\begin{cases}
 			1 \quad \text{if } w=\max_{i\in \{1, 2, 3 \}}[t-t_i]^+ , \\
 			0 \quad \text{else},
 		\end{cases} \nonumber \\
 		\text{ where }
 		[x]^+
 		= &\, 
 		\begin{cases}
 			x \quad \text{if } x\ge 0 , \\
 			0 \quad \text{else}.
 		\end{cases}
    \end{align}
    The vaccination status $v$ also depends deterministically on the vaccination times $t_1, t_2, t_3$:
    \begin{equation}
 		P(v|t_1, t_2, t_3, t)
 		=
 		\begin{cases}
 			1 \quad \text{if } v=\max_{i\in \{0, 1, 2, 3 \}} \left(i \cdot \mathrm{sgn}(t-t_i+1) \right) , \\
 			0 \quad \text{else},
 		\end{cases}
 	\end{equation}
 	where $\mathrm{sgn}$ is the sign function.
 	These deterministic relationships can be used to eliminate the corresponding conditionals from the target function:
 	\begin{linenomath*}  %
     	\begin{align}
        	&\ s(\tilde \pi = \tilde P(T_1, T_2, T_3|A)) \nonumber \\
        	&=
        	D \sum_a P(a) 
            \sum_{t=1}^M \nonumber\\
        	& \quad \times\left[
        	\sum_{t_1=t+1}^{M+1} \sum_{t_2=t+1}^{M+1} \sum_{t_3=t+1}^{M+1} \tilde P(t_1, t_2, t_3|a) \
        	P(S{=}1|t, a, v{=}0, w{=}0) \right. \nonumber\\ 
        	& \quad +
        	\sum_{t_1=1}^{t} \sum_{t_2=t+1}^{M+1} \sum_{t_3=t+1}^{M+1} \tilde P(t_1, t_2, t_3|a) \
        	P(S{=}1|t, a, v{=}1, w{=}(t-t_1)) \nonumber\\
        	& \quad +
        	\sum_{t_1=1}^{t} \sum_{t_2=1}^{t} \sum_{t_3=t+1}^{M+1} \tilde P(t_1, t_2, t_3|a) \
        	P(S{=}1|t, a, v{=}2, w{=}(t-t_2)) \nonumber\\
        	& \quad \left. +
        	\sum_{t_1=1}^{t} \sum_{t_2=1}^{t} \sum_{t_3=1}^{t} \tilde P(t_1, t_2, t_3|a) \
        	P(S{=}1|t, a, v{=}3, w{=}(t-t_3)) \right].
        \end{align}
    \end{linenomath*}
    Substituting the factorisation for the severity mechanism $P(S{=}1|v, a, t, w)$ we get:
    \begin{linenomath*}  %
        \begin{align}
        	&\ s(\tilde \pi = \tilde P(T_1, T_2, T_3|A)) \nonumber \\
        	&=
        	D \sum_a P(a) 
            \sum_{t=1}^M f^0(t) f^1_{\tilde \pi}(a, t) \nonumber\\
        	& \quad \times  \left[
        	\sum_{t_1=t+1}^{M+1} \sum_{t_2=t+1}^{M+1} \sum_{t_3=t+1}^{M+1} \tilde P(t_1, t_2, t_3|a) \,
        	\, g(0, a) \right. \nonumber\\ 
        	 & \quad +
        	\sum_{t_1=1}^{t} \sum_{t_2=t+1}^{M+1} \sum_{t_3=t+1}^{M+1} \tilde P(t_1, t_2, t_3|a) \,
        	\, g(1, a) \, h^1(t-t_1)\nonumber\\
        	 & \quad +
        	\sum_{t_1=1}^{t} \sum_{t_2=1}^{t} \sum_{t_3=t+1}^{M+1} \tilde P(t_1, t_2, t_3|a) \,
        	\, g(2, a) \, h^2(t-t_2)\nonumber\\
        	 & \quad \left. +
        	\sum_{t_1=1}^{t} \sum_{t_2=1}^{t} \sum_{t_3=1}^{t} \tilde P(t_1, t_2, t_3|a) \,
        	\, g(3, a) \, h^3(t-t_3) \right].
        \end{align}
    \end{linenomath*}

    \newpage
    \section{Results assuming other mixing factors} %
    \label{app:results_assuming_other_mixing_factors}
    
    \begin{figure}[htb]
        \centering
        \includegraphics[width=\singleColWidth]{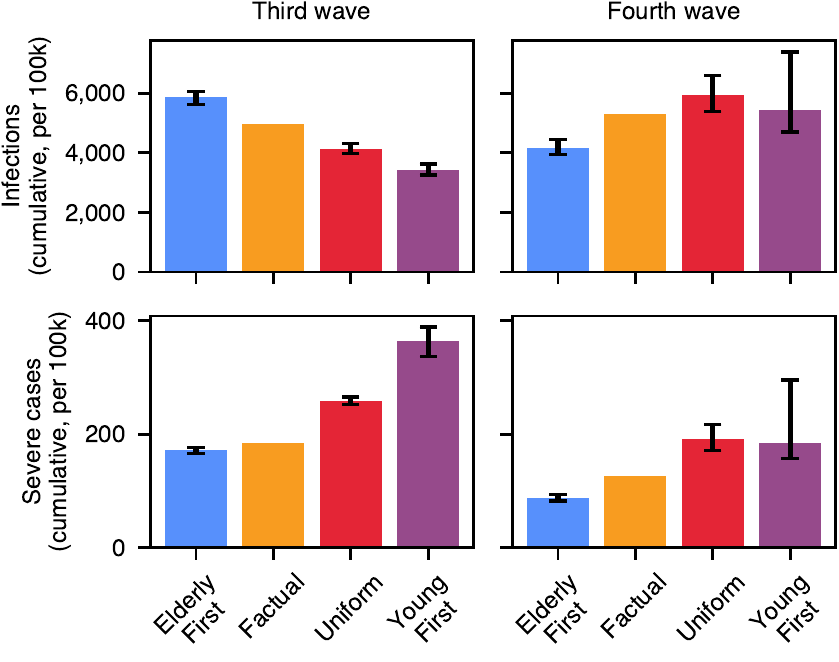}
        \caption{
        \looseness=-1
        \textbf{Cumulative incidences of infections (top row) and severe cases (bottom row) for the two infection waves in 2021 under the factual and counterfactual vaccine allocation strategies, assuming a contact mixing factor of $0.7$.}
        For the third wave we sum all cases from \DTMdisplaydate{2020}{12}{20}{-1} to \DTMdisplaydate{2021}{04}{11}{-1}; for the fourth wave from \DTMdisplaydate{2021}{06}{20}{-1} to \DTMdisplaydate{2021}{11}{07}{-1}.
        The whiskers show the $95\%$ credible intervals.
        }
        \label{fig:policy_exp_cumulative_c70}
    \end{figure}
    
    \begin{figure}[htb]
        \centering
        \includegraphics[width=\singleColWidth]{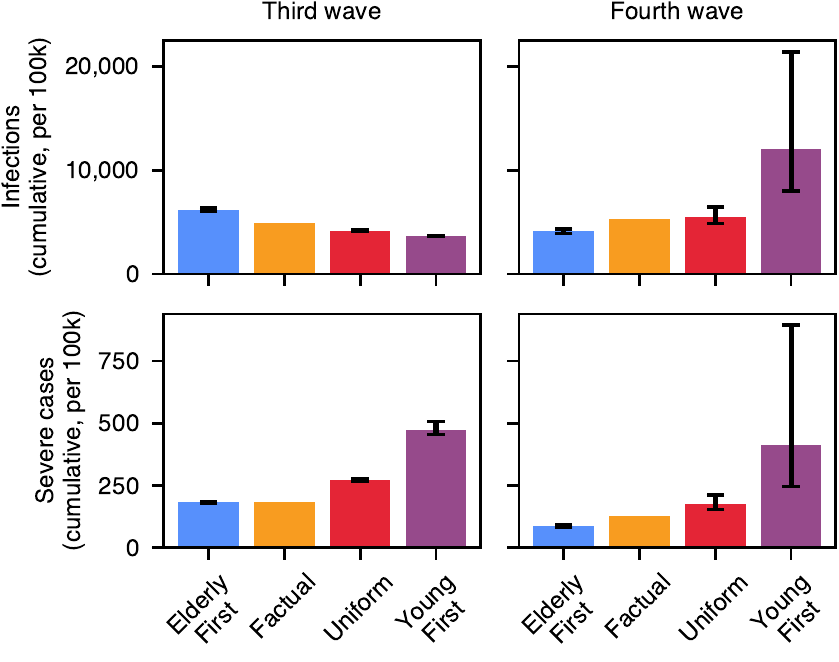}
        \caption{
        \looseness=-1
        \textbf{Cumulative incidences of infections (top row) and severe cases (bottom row) for the two infection waves in 2021 under the factual and counterfactual vaccine allocation strategies, assuming a contact mixing factor of $0.9$.}
        For the third wave we sum all cases from \DTMdisplaydate{2020}{12}{20}{-1} to \DTMdisplaydate{2021}{04}{11}{-1}; for the fourth wave from \DTMdisplaydate{2021}{06}{20}{-1} to \DTMdisplaydate{2021}{11}{07}{-1}.
    The whiskers show the $95\%$ credible intervals.
        }
        \label{fig:policy_exp_cumulative_c90}
    \end{figure}
    
    \begin{figure}
        \centering
        \includegraphics[width=\singleColWidth]{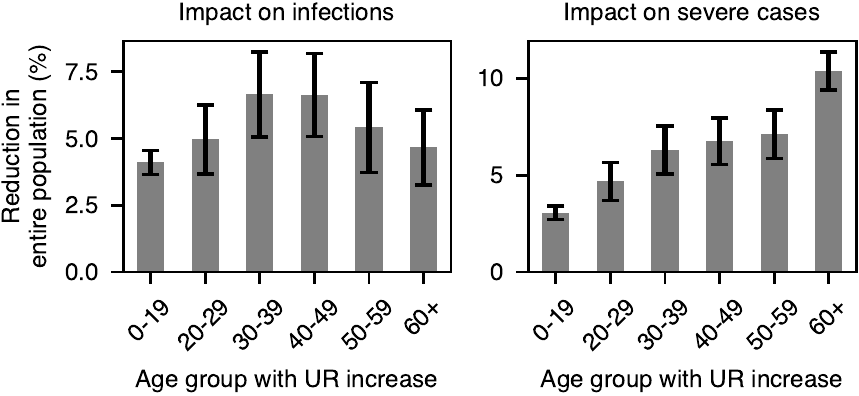}
        \caption{
            \textbf{Impact of increasing vaccine uptake rate (UR) in a given age group on severe cases, assuming a contact mixing factor of $0.7$.}
            In each scenario, the vaccine uptake rate is increased in a given age group by a fixed number corresponding to $0.6 \%$ of the population being motivated to get vaccinated.
            We assume that the change comes from originally unvaccinated individuals who are persuaded to receive three doses.
            The plots show the impact on cumulative infections (left) and severe cases (right) in the entire population---not just in the age group in which the UR was increased.
            We consider cases from \DTMdisplaydate{2020}{12}{20}{-1} to \DTMdisplaydate{2021}{12}{25}{-1}.
            The whiskers show the $95\%$ credible intervals.
            }
        \label{fig:acceptance_exp_overview_c70}
    \end{figure}
    
    \begin{figure}
        \centering
        \includegraphics[width=\singleColWidth]{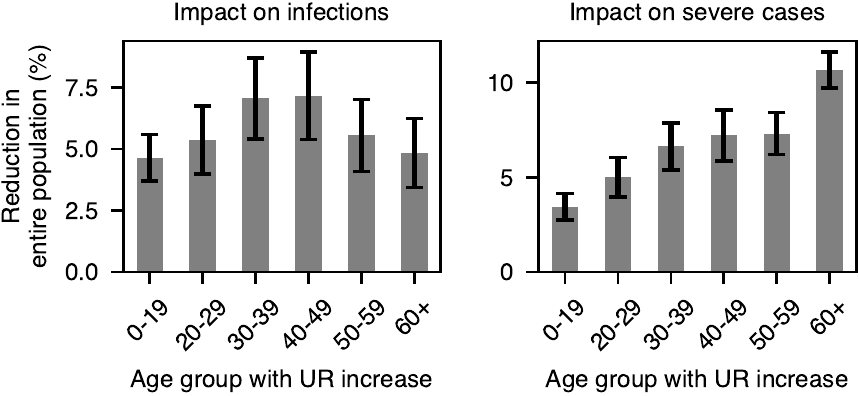}
        \caption{
            \textbf{Impact of increasing vaccine uptake rate (UR) in a given age group on severe cases, assuming a contact mixing factor of $0.9$.}
            In each scenario, the vaccine uptake rate is increased in a given age group by a fixed number corresponding to $0.6 \%$ of the population being motivated to get vaccinated.
            We assume that the change comes from originally unvaccinated individuals who are persuaded to receive three doses.
            The plots show the impact on cumulative infections (left) and severe cases (right) in the entire population---not just in the age group in which the UR was increased.
            We consider cases from \DTMdisplaydate{2020}{12}{20}{-1} to \DTMdisplaydate{2021}{12}{25}{-1}.
            The whiskers show the $95\%$ credible intervals.
            }
        \label{fig:acceptance_exp_overview_c90}
    \end{figure}
    \FloatBarrier

\end{appendix}

\end{document}